\documentclass[onecolumn,showpacs,superscriptaddress,11pt]{revtex4-2}
\usepackage{amsmath,amssymb,graphicx}
\usepackage{bm, amsmath,amssymb,graphicx, subfigure, braket, float}
\usepackage{xcolor}
\usepackage{graphicx}

\usepackage[left=0.5in, right=0.5in, top=0.5in, bottom=0.5in]{geometry}
\usepackage{xcolor}
\usepackage{graphicx}
\usepackage[compact]{titlesec}

\usepackage{xcolor}
\usepackage{color}
\usepackage{hyperref}
\hypersetup{colorlinks=true,citecolor=blue,filecolor=blue,linkcolor=blue,urlcolor=blue}

\newcommand{\era}{\end{array}}
\newcommand{\beq}{\begin{equation}}
\newcommand{\eeq}{\end{equation}}
\newcommand{\beqar}{\begin{eqnarray}}
\newcommand{\eeqar}{\end{eqnarray}}

\def\BC{\bb C}
\def\_\BC{\bbi C}

\def\( {\left(}
\def\) {\right)}
\def\[ {\left[}
\def\] {\right]}
\def\no2 {{\textstyle{n\over 2}}}

\newcommand{\lb}{\label}

\def\keywordname{{\bfseries \emph{Keywords}}}%
\def\keywords#1{\par\addvspace\medskipamount{\rightskip=0pt plus1cm
\def\and{\ifhmode\unskip\nobreak\fi\ $\cdot$
}\noindent\keywordname\enspace\ignorespaces#1\par}}

\begin{document}
\title{Probing a hybrid channel for the dynamics of non-local features}
\author{Atta ur Rahman$^{1}$,  Ma-Cheng Yang$^{1}$, S. M. Zangi$^{2}$, Cong-Feng Qiao \footnote{\textcolor{blue}{qiaocf@ucas.ac.cn}}}
\address{\small School of Physics, University of Chinese Academy of Science, Yuquan Road 19A,  Beijing 100049, China\\
$^{2}$School of Physics and Astronomy, Yunnan University, P.O. Box 650500 Kunming, China}

\begin{abstract}
\bigskip
Effective information transmission is a central element in quantum information protocols, but the quest for optimal efficiency in channels with symmetrical characteristics remains a prominent challenge in quantum information science. In light of this challenge, we introduce a hybrid channel that encompasses thermal, magnetic, and local components, each simultaneously endowed with characteristics that enhance and diminish quantum correlations. To investigate the symmetry of this hybrid channel, we explore the quantum correlations of a simple two-qubit Heisenberg spin state, quantified using measures such as negativity, $\ell_1$-norm coherence, entropic uncertainty, and entropy functions. Our findings reveal that the hybrid channel can be adeptly tailored to preserve quantum correlations, surpassing the capabilities of its individual components. We also identify optimal parameterizations to attain maximum entanglement from mixed-entangled/separable states, even in the presence of local dephasing. Notably, various parameters and quantum features, including non-Markovianity, exhibit distinct behaviors in the context of this hybrid channel. Ultimately, we discuss potential experimental applications of this configuration.

	\end{abstract}
	\renewcommand{\baselinestretch}{1.3} \topmargin=-1.5cm \textheight=23 cm
	\textwidth=24cm
\maketitle
\section{Introduction}
\par
The transmission of information in quantum information processing is done using various types of communicating means which include local and non-local channels \cite{42}. The designing of such transmitting channels is important because the resourcefulness of the quantum systems may be lost if the inclusive channels have disorders or no optimal characteristics. In the previous era, various types of channels with various characteristics have been treated separately. However, this would be far from reality, as the characteristics of these various channels can be found at the same place simultaneously, for example, the thermal, magnetic, and dephasing effects. In this regard, the thermal interaction picture of the channels has been studied using the concept of Gibb's density matrix operation \cite{a11}. Besides this, magnetic fields have also been used to demonstrate the quantum correlations dynamics explicitly \cite{a12}. Most importantly, the local channels are successfully used for decades for the transmission of information \cite{45}. On the contrary, non-local channels have been investigated and a lot of efforts for their practical empowerment have been made recently \cite{46}. Compared to the non-local ones, the classical ones have the advantage of being easily implemented, as they do not need any complex design. Pure classical channels, nonetheless, are influenced by certain types of flaws and disorders, such as the impact of surface charge carriers \cite{47}, electronic currents \cite{48}, thermal fluctuations \cite{49, 50} and so forth. One of these disorders causes static noise, which will be considered to influence classical channels in the current situation \cite{51}. Previously, static noise has been widely studied in independent classical channels and has been found to completely degrade quantum correlations \cite{49, 50, 51}.
\par
We are planning to demonstrate the effectiveness of a complex mixed channel, deployed for information transmission, therefore, we will consider simple two-qubit correlations to evaluate our goal. In the emerging discipline of quantum information science, non-classical correlations are now the primary priority of foundational scientific study \cite{1}. A crucial non-local resource known as quantum entanglement explains a particular kind of non-classical correlation among subsystems of a quantum composite state \cite{2}. When a group of particles interacts spatially such that the quantum states of each particle are mutually exclusive, this is referred to as quantum entanglement. This non-local correlation causes measurements on one part of an entangled pairing and immediately affects the results of measurements on the other. Quantum computers can perform computations that are not conceivable on classical computers because of the advantage that non-local systems can exist in multiple states simultaneously \cite{3}. Researchers are actively developing a quantum internet and quantum encrypted communications using entanglement \cite{6}, which would allow for new varieties of telescopes \cite{7}, sensors \cite{8}, and ultra-secure broadcasting \cite{9}.  In addition, the information encoded in qubits can be processed more quickly and with less processing power because of the quantum entanglement phenomenon. Recently, it has been demonstrated that quantum coherence and entanglement, two fundamentally dissimilar aspects of quantum theory, are functionally equivalent and made a variety of quantum technologies possible, and therefore, both would remain part of the current study \cite{15}.
\par
The uncertainty principle holds that the results of simultaneous measurements of non-commuting observables commonly have an intrinsic lower bound on the uncertainty \cite{18}. This assumption both illustrates how the classical and quantum realms differ and provides the rationale for the ambiguity of quantum mechanics. In essence, it states that you cannot begin preparing a quantum particle whose location and momentum are predictable with certainty, simultaneously \cite{20}. Kennard and Robertson revisited the principle of uncertainty after it had taken the form of a standard deviation and was based on the combined variance of two observables \cite{21}. For example, for two observables $\mathcal{X}$ and $\mathcal{Y}$, the associated uncertainty relation was defined as $\Delta\mathcal{X}\Delta\mathcal{Y} \geq 1/2 \vert \langle [ \mathcal{X}, \mathcal{Y}] \rangle \vert $. Furthermore, Deutsch established the well-known version of the entropic uncertainty (EU) relation in 1983 and proposed the uncertainty principle in terms of Shannon entropy \cite{23}. Afterward, Kraus first proposed another simplified version of the EU relation which was verified in 1988 by Maassen and Uffink \cite{24, 25}. Besides, for different quantum systems and situations, various EU relations have been established The study of the EU will provide an important aspect of the current hybrid channel to induce uncertainty in the process of quantum information processing \cite{23d1,23d2,23d3}.
\par

\par
In this work, we consider examining a two-qubit spin state influenced by a thermal and magnetic field \cite{55}. The spin system is further assumed to be characterized by the Dzyaloshinskii-Moriya interaction (DM), Kaplan, Shekhtman, Entin-Wohlman, and Aharony (KSEA) and anisotropy parameters interactions \cite{56, 57}. The DM interaction, an antisymmetric exchange interaction that controls chiral spin configurations, is brought on by inversion symmetry breaking in non-centrosymmetric crystal lattice interfaces \cite{58}. On the other hand, the KSEA interaction is caused by the symmetric helical interaction \cite{57}. Besides, the generated quantum correlations need interaction with external channels to be transmitted. As it is known that classical channels are highly destructive for quantum correlations, therefore, we become motivated to investigate the impact of the joint implications of the external thermal-magnetic-classical channel (TMCC) on quantum correlation dynamics. The reason behind designing such a configuration is to devise a reliable and physically constructional transmitting channel which we believe will be helpful in quantum correlations preservation during communication using non-local systems. The entanglement measurement in the spin state is done by using negativity (NG), which is one of the well-known bipartite entanglement monotones \cite{59}. As stated before that entanglement and coherence are two fundamental correlations of quantum systems. Motivated by this, we also analyze the dynamics of coherence using $\ell_1$-norm of coherence in the spin system \cite{60}. Quantum systems are always influenced by uncertainty, therefore, we also consider assessing the degree of uncertainty between different observables of the system using EU relations \cite{61}. Besides uncertainty, mixedness disorder is a commonly found phenomenon where most of the quantum correlations in quantum systems are lost because of its emergence. The degree of disorder in the spin state when subjected to the hybrid channel with thermal, magnetic, and local dephasing parts will be computed utilizing linear entropy (EN) \cite{62}. One of the basic motivations behind the simultaneous measurement of entanglement, coherence, entropic uncertainty, and mixedness disorder is to find out the reaction of the assumed hybrid channel towards different non-classical characteristics.
\par

This work is presented as follows: In Sec. \ref{model}, we provide the details of the physical model of the assumed system and channel, along with an introduction to the measurement of two-qubit correlations. Sec. \ref{results} gives a detailed analysis of the results obtained, and in Sec. \ref{conclusion}, we summarize this work.

\section{Physical model}\label{model}
\subsection{Thermal and magnetic interaction}
We assume a two spin-$1/2$ XXZ type Heisenberg system when exposed to an external homogeneous magnetic field characterized by the DM ($D_z$) and KSEA $(K_z)$ coupling interaction oriented along the $z$-axis. This configuration has the Hamiltonian model given by \cite{36p}
\begin{equation}
\textbf{H}=\Delta_z \Big( S_{1}^{z}S_{2}^{z} \Big)+  D_z\Big(S_{1}^{x}S_{2}^{y}-S_{1}^{y}S_{2}^{x}\Big)+J\Big(S_{1}^{x}S_{2}^{x}+ S_{1}^{y}S_{2}^{y}\Big)+ K_z \Big(S_{1}^{x}S_{2}^{y}+S_{1}^{y}S_{2}^{x}\Big)+ B\Big(S_1^z + S_2^z \Big), \label{Hmatrix}
\end{equation}
\begin{figure}
\includegraphics[scale=0.06]{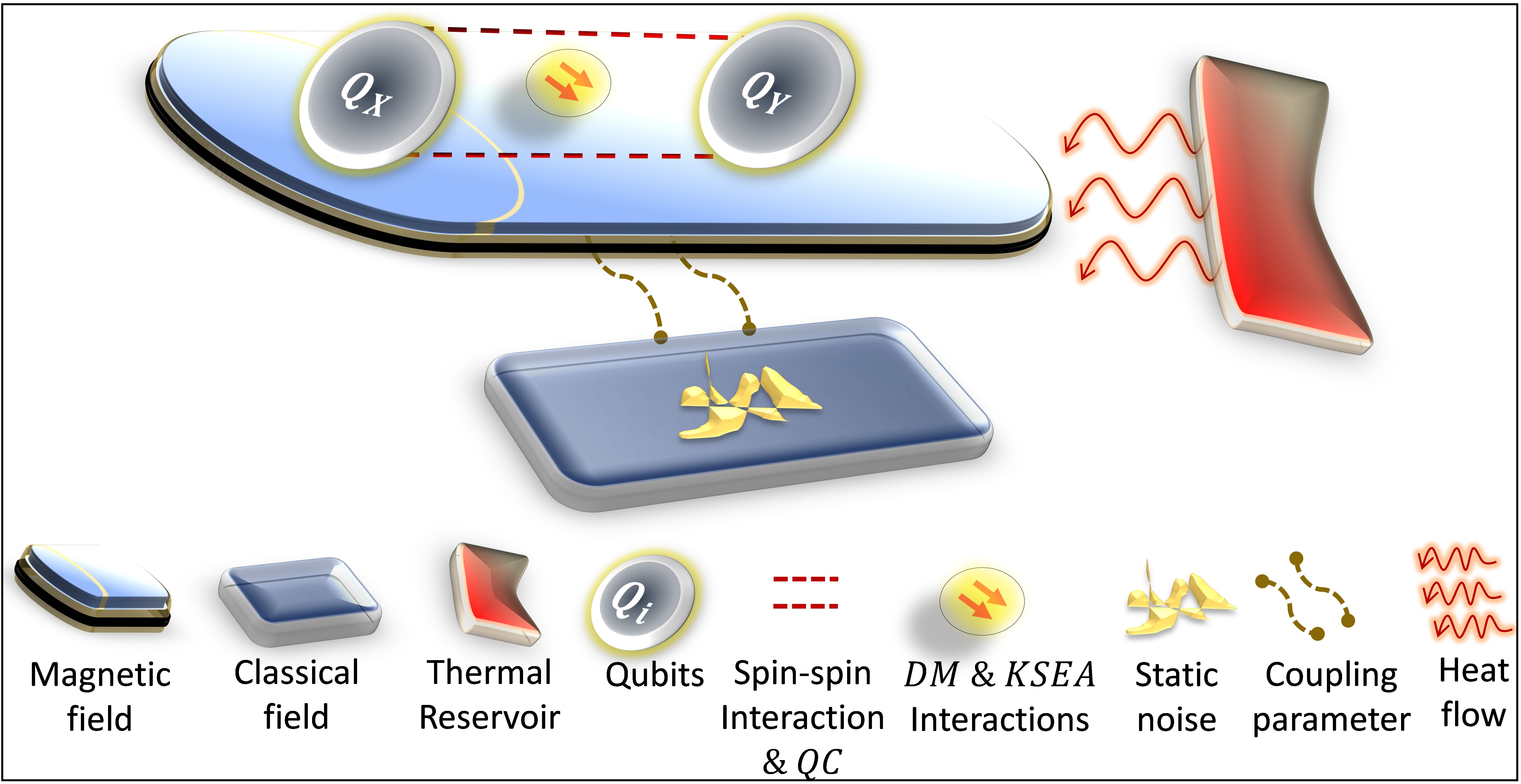}
\caption{Shows the physical model of the hybrid channel with thermal, magnetic, and classical dephasing parts controlled by static noise employed for the dynamics of two-qubit Heisenberg spin state characterized by various parameters, such as spin-spin, DM, and KSEA interaction.}
\end{figure}
where $S_{f}^{g}$ with $(f=\{ 1,~2\},~g=\{x,~y,~z\})$ are the spin-$1/2$ Pauli matrices of the spin $f$, $\Delta_z $ is the real anisotropy coupling constant describing the symmetric exchange spin-spin interaction in $z$-direction, $D_z$ is the DM interaction strength regulating the spin-orbit antisymmetric coupling, $J$ is the Heisenberg exchange interaction between the spin particles. Note that for $J > 0$($J < 0$) correspond to the antiferromagnetic(ferromagnetic) interaction between the spins sites, respectively. Besides, $K_z$ represents the KSEA interaction strength oriented along $z-$axis responsible for symmetric spin-orbit coupling while $B$ is the homogeneous part of the considered magnetic field. The Hamiltonian  expression given in Eq. \eqref{Hmatrix} in the computational basis $\lbrace \vert 00 \rangle, \vert 01 \rangle, \vert 10 \rangle, \vert 11 \rangle \rbrace$ takes the form

\begin{equation}\label{Hmatrix}
  \textbf{H}=
	 \left[
      \begin{array}{cccc}
      2B+\Delta_z &0&0&-2 i K_z\\[0.2cm]
      0&-\Delta_z  & 2iD_z+2 J&0\\[0.2cm]
      0&-2iD_z+2J &-\Delta_z &0\\[0.2cm]
      2 i K_z&0&0&-2B+\Delta_z
      \end{array}
   \right].
\end{equation}
\par
The  corresponding eigenvectors of the Hamiltonian $\textbf{H}$ are

	\begin{align*}
 \vert \Theta_{1^+,4^-}\rangle&=\sqrt{\frac{\Lambda \pm 2B}{2\Lambda}}\left(|00\rangle \pm \frac{ 2iK_z}{{\Lambda \pm 2 B}}\vert11\rangle\right),&
\vert \Theta_{2^+,3^-}\rangle &=\frac{1}{\sqrt{2}} \left(|01\rangle \pm \frac{2 J \pm 2iD_z}{\upsilon}\vert10\rangle\right),&\end{align*}

with $\Lambda=\sqrt{4B^2+4 K_{z}^{2}}$ and $\upsilon=\sqrt{4J^2+4D_{z}^{2}}$. Besides, the eigenvalues of Eq. \eqref{Hmatrix} are $\lb{eq1} E_{1^+,4^-}= \Delta_z \pm \Lambda$, $\lb{eq2} E_2= -\Delta_z + \upsilon$,
\newline
 Using Gibbs's density operator, the thermal state density matrix for the Hamiltonian given in Eq. \eqref{Hmatrix} in equilibrium with a thermal reservoir at temperature $T$ has the form
\begin{eqnarray}
\rho(0, T) = \frac{1}{{Z}} \exp \Big\{\frac{-\textbf{H}}{k_B T} \Big \}=\left[\begin{array}{llll}
\rho_{11} & 0 & 0 & \rho_{14} \\[0.2cm]
0 & \rho_{22} & \rho_{23} & 0 \\[0.2cm]
0 & \rho_{32}^{\ast} & \rho_{33} & 0 \\[0.2cm]
\rho_{41}^{\ast} & 0 & 0 & \rho_{44}
\end{array}\right],\label{state}
\end{eqnarray}
where $Z= {\rm Tr}[ \exp\{\frac{-\textbf{H}}{T}\}] = \sum^{4}_{i=1} \exp \{\frac{-E_i}{T}\}$ is the partition function with $k_B=1$ is the Boltzmann's constant. 
The corresponding entries are given by
\begin{align}
\rho_{11}&=\frac{1}{Z}e^{-\frac{\Delta_z}{T}} \left(\cosh (\varphi)-\frac{B \sinh (\varphi)}{\sqrt{B^2+K_z^2}}\right),& 
\rho_{14}&=\rho_{41}^*=\frac{1}{Z \sqrt{B^2+K_z^2}}i K_z e^{-\frac{\Delta_z}{T}} \sinh (\varphi),&\nonumber\\
\rho_{22}&=\rho_{33}=\frac{1}{Z}e^{\Delta_z}/T \cosh (\varpi),& 
\rho_{23}&=\rho_{32}^*=\frac{1}{Z \sqrt{D_z^2+J^2}}(-J-i D_z) e^{\Delta_z}/T \sinh (\varpi),&\nonumber\\
\rho_{44}&=\frac{1}{Z}e^{-\frac{\Delta_z}{T}} \left(\frac{B \sinh (\varphi)}{\sqrt{B^2+K_z^2}}+\cosh (\varphi)\right), \label{rho_0} &
\end{align}
where $\varphi=\frac{1}{T}(2 \sqrt{B^2+K_z^2})$, $\varpi=\frac{1}{T}(2 \sqrt{\text{Dz}^2+J^2})$ and partition function $Z=2 e^{-\frac{\Delta_z}{T}} \left(e^{\frac{2 \Delta_z}{T}} \cosh (\varpi)+\cosh (\varphi)\right).$
\par
The eigenvalues of the two-qubit density matrix $\rho(0,T)$ are given by
	\begin{align}
   	E_{1^+,2^-} =&\frac{1}{2} \left(\pm \sqrt{4 \rho_{14} \rho_{14}^*+(\rho_{11}-\rho_{44})^2}+\rho_{11}+\rho_{44}\right),&
   	E_{3^+,4^-} =&\frac{1}{2} \left(\pm\sqrt{4 \rho_{23} \rho_{23}^*+(\rho_{22}-\rho_{33})^2}+\rho_{22}+\rho_{33}\right).
   		\end{align}

\subsection{The exposure to a classical channel}
Here we provide the exposure of the two-qubit spin state to a common classical environment driven by static noise. In the present case, the Hamiltonian, which governs the current physical model is written as \cite{63}
\begin{equation}
{\rm \bf H}_{XY}={\rm \bf H}_X \otimes {\rm I}_Y+{\rm I}_X \otimes{\rm \bf H}_Y,~~\text{with}~~{\rm \bf H}_\mathcal{P}=\left[
\begin{array}{cc}
 \Delta_\mathcal{P} \lambda +\epsilon  & 0 \\
 0 & \epsilon - \Delta_\mathcal{P} \lambda  \\
\end{array}
\right],
\label{hmm}
\end{equation}
where $ {\rm \bf H}_{\mathcal{P}}~(\mathcal{P}=X, Y)$ denotes the Hamiltonian state of the sub-system $\mathcal{P}$, $\epsilon$ is the equal energy splitting between the sub-systems, $I$ is the $2 \times 2$ identity matrix, $\lambda$ is the coupling constant, $\Delta_\mathcal{P}$ regulates the stochastic behavior of the classical field and is flipping between $\pm1$, and $S^z$ is the spin-Pauli matrix.
\par
For the time evolution of the system, we use the following time-unitary operator
where
\begin{equation}
U_{XY}(t)=\exp \Big\{-i \int^{t}_{t_0}H(z)dz \Big\}=\left[
\begin{array}{cccc}
 U_{11} & U_{12} & U_{13} & U_{14} \\[0.2cm]
 U_{12} & U_{11} & U_{14} & U_{13} \\[0.2cm]
 U_{13} & U_{14} & U_{11} & U_{12} \\[0.2cm]
 U_{14} & U_{13} & U_{12} & U_{11} \\
\end{array}
\right],
\end{equation}
which is the time-unitary matrix with $\hbar=1$ and the matrix entries are
\begin{align*}
U_{11}=&\exp \{ -2 i t \epsilon \} \cos (\Delta_X \lambda  t) \cos (\Delta_Y \lambda  t),&
 U_{12}=&-i \exp \{-2 i t \epsilon \} \cos (\Delta_X \lambda  t) \sin (\Delta_Y \lambda  t),&\\
U_{13}=&-i \exp \{-2 i t \epsilon \} \sin (\Delta_X \lambda  t) \cos (\Delta_Y \lambda  t),&
 U_{14}=&-\exp \{-2 i t \epsilon \} \sin (\Delta_X \lambda  t) \sin (\Delta_Y \lambda  t).&
\end{align*}
To obtain the time-evolved state of the system for the initial thermal state density matrix  $\rho(0, T)$ given in Eq. \eqref{rho_0} when exposed to identical channel i.e. $\Delta_X=\Delta_Y$, we use \cite{63}
\begin{equation}
\rho_{X, Y}(t, T)=U_{XY}(t)\rho(0, T) U_{XY}(t)^{\dagger} \rightarrow\rho(t, T)=U_{XX}(t)\rho(0, T)U_{XX}(t)^{\dagger},\label{time evolved density matrix}
\end{equation}
and explicitly, the above equation takes the following shape:

\begin{align}
\rho(t, T)=\left[
\begin{array}{cccc}
 \rho_{11} & 0 & 0 &  e^{-4 i \Delta_X \lambda  t} \rho_{14}\\[0.2cm]
 0 & \rho_{22} & \rho_{23} & 0 \\[0.2cm]
 0 & \rho_{32}^* & \rho_{33} & 0 \\[0.2cm]
 \rho_{41}^* e^{4 i \Delta_X \lambda  t} & 0 & 0 & \rho_{44} \\
\end{array}
\right].\label{rho-t}
\end{align}
\subsubsection{The influence of classical static noise disorder}
Next, we provide the application and influence of the static noise on the time-evolved state of the spin system. In this regard, static noise is primarily characterized by $ \Delta_Q $, namely, the disorder parameter. This noise has the probability distribution function $ \mathcal{O}(\delta)=1/\Delta_Q$ and exhibits the range $ |\delta-\delta_o| \leq \Delta_Q / 2 $ where $ \delta_o $ denotes the mean value of the probability distribution function \cite{51s}. The static noise auto-correlation expression is written as $\langle \delta \Delta(t) \Delta(0) \rangle=\Delta^{2}_Q /12$. As a result, this kind of noise tends to have a characteristic duration that is significantly longer than the coupling between the system and its environment. To evaluate the impact of the static noise on the dynamics of the spin state, the time-evolved state density matrix is averaged over all possible noise configurations. Therefore, we integrate the matrix in Eq. \eqref{rho-t}  between $r^{+}=\delta_o-\Delta_Q/2$ and $r^{-}=\delta_o+\Delta_Q/2$ as \cite{51}
\begin{equation}
\rho_{st}(t, T)=\int^{r^{+}}_{r^{-}}\frac{1}{\Delta_Q}\rho(t, T) d\Delta_X=\left[
\begin{array}{cccc}
\rho_{11} & 0 & 0 & \hat{\rho}_{14} \\[0.2cm]
 0 & \rho_{22} & \rho_{23} & 0 \\[0.2cm]
 0 & \rho_{23}^* & \rho_{33} & 0 \\[0.2cm]
 \hat{\rho}_{41}^* & 0 & 0 & \rho_{44} \\
\end{array}
\right], \label{final rho}
\end{equation}
where $\rho_{ii}(ii=11,22,33,44,23)$
remains the same as given in Eq. \eqref{rho_0}. However,
$\hat{\rho}_{14}=\hat{\rho}_{41}^*=\frac{\rho_{14} e^{-4 i \delta_o \lambda  t} \sin (2 \Delta_Q \lambda  t)}{2 \Delta_Q \lambda  t}.$\par
Note that the final matrix given in Eq. \eqref{final rho} represents the two-spin system simultaneously influenced by a thermal, magnetic, and classical channel. Note that the off-diagonal matrix elements are included now with dephasing terms hence, coherence loss of the systems is included. Therefore, the loss of entanglement, rise of the uncertainty, and mixedness in the state with time would give us a realistic model in comparison, if one would only consider the case of magnetic and thermal interaction.

\subsection{Quantum criteria quantifiers}
\subsubsection{Bipartite Negativity}
Multiple quantitative factors have been developed that demonstrate the level of entanglement in a quantum state. For example, negativity (${\rm NG}$) has been investigated as a valuable and calculable witness to entanglement for any pure and mixed states. The Peres-Horodecki separability criterion serves as the foundation for this criterion. On a rescaled scale, {\rm NG} for a statistical ensemble $\rho_{XY}$ is defined as \cite{63a}
\begin{align}
{\rm NG}=2\sum_i|\mathcal{Q}_i|,\label{NG}
\end{align}
where $\mathcal{Q}_i$ are the negative eigenvalue(s) of the partially transposed density matrix $\rho_{XY}^{T_X}$ with respect to sub-system $X$. The state is maximally correlated if $NG=1$ while for $NG=0$, the state will become separable \cite{63}.

\subsubsection{The entropic uncertainty measure}
Consider Bob and Alice to be the two users. Alice receives a qubit in the desired quantum state created by Bob. Alice must now select one of the two measurements and notify Bob of her selection. We can now lower the result's uncertainty by using Bob's measurement data \cite{21}. The uncertainty standard deviation for two observables $\mathcal{X}$ and $\mathcal{Y}$, can be represented as: $\Delta \mathcal{X} \Delta \mathcal{Y} \geq \frac{1}{2} |\langle[\mathcal{X}, ~\mathcal{Y}] \rangle|$ \cite{23, 23d1, 23d2, 23d3}.
\par
Instead of utilizing the standard deviation to describe uncertainty, Deutsch suggested the entropic uncertainty relation for every pair of observables, as shown in the above equation. Based on Deutsch's method, Maassen and Uffink created a tighter entropic uncertainty formulation, which can be represented as $S(\mathcal{X})+S(\mathcal{Y})\geq \log_2c(\mathcal{X};\mathcal{Y}),$ \cite{64} where $S(K)(K=\mathcal{X},\mathcal{Y})$ is known as the Shannon entropy denoting the probability distribution of measuring the observable $K$ while $c(\mathcal{X};\mathcal{Y})=\max_{a,~b}\vert \langle \psi \vert \phi \rangle \vert^2$ is the maximal overlap between the eigenvectors $\psi$ and $\phi$ of the two non-degenerate observables. \par
A new definition presented by Renes et al. and Berta et al. considered the entropic uncertainity relations in composite systems and proposed a quantum-memory-assisted entropic uncertainty relation \cite{65, 65b}. Therefore, the quantum-memory assisted entropic uncertainty regarding system $A$ is reproduced corresponding to q quantum memory $B$ with a tighter bound, and has the form
\begin{align}
S(\mathcal{X}\vert B) + S(\mathcal{Y}\vert B) \geq S(A \vert B)-\log_2c(\mathcal{X};\mathcal{Y})),\label{ULR}
\end{align}
where $S(\mathcal{X} \vert B)=S(\rho_{\mathcal{X} B})-S(\rho_B)$ is the conditional von Neumann entropy of the system after the unilateral measurement is applied on $\mathcal{X}$ on subsystem $A$. Note that the same procedure can be carried our for $\mathcal{Y}$ too. The associated  post-measurement state now can be written as
\begin{equation}
\rho_{\mathcal{X} B}=\sum_i(\vert \psi \rangle \langle\psi \vert \otimes I)\rho_{AB}(\vert \psi \rangle \langle\psi \vert \otimes I).
\end{equation}
The conditional entropy $S(A|B)$ computes the correlation between subsystem $B$ owned by its memory system $A$.
\par
Eq. \eqref{ULR} comprises the entropic uncertainty (EU) on the left-hand side and entropic uncertainty lower bound on the right-hand side. In the current case, we only emphasize measuring the left-hand side of the uncertainty relations.
\subsubsection{$\ell_1$-norm of coherence}
Different metrics, including the distance metric \cite{66}, the relative entropy of coherence \cite{31} and purity \cite{49} can be used to evaluate the coherence. The $\ell_1$-norm is a reliable coherence monotone and a coherence criterion \cite{31}. Besides, the $\ell_1$-norm coherence for a two-qubit state  $\rho=\sum_{X, Y}\rho_{XY} \vert X\rangle \langle Y \vert$ is the sum of all the off-diagonal entries as \cite{67}
\begin{equation}
LC(t)=\sum_{X \neq Y}|\rho_{XY}|, \label{L1norm}
\end{equation}
where $|\rho_{XY}|$ is the absolute value of the elements of the density matrix $\rho_{XY}$.
\subsubsection{Linear entropy}
Linear entropy, a simple to evaluate scalar field, is used to measure the mixedness of quantum states. For example, by approximating the $\log_{2}\rho_{XY}$ of $S(\rho_{XY})$ with the first-order term i.e., $\rho_{XY}-1$ in the Mercator series given in Ref. \cite{49}, one may get
\begin{align}
EN(t)=&-\operatorname{Tr}[\rho_{XY}\log_2\rho_{XY}]. \label{LT}
\end{align}

For the next-to-last equality, the density matrix's unit trace feature ($\operatorname{Tr}[\rho_{XY}]=1$) is used.

\section{Results}\label{results}
This section is devoted to exploring the numerical results obtained for the dynamics of the two-qubit spin state influenced by an external TMCC configuration. Additionally, the impact of various parameters such as anisotropy, spin coupling strength, DM and KSEA interactions as well as the external noise parameters of the static noise is studied. Utilizing $NG$, $EU$, $LC$, and $EN$ functions given in Eqs. \eqref{NG}, \eqref{ULR}, \eqref{L1norm}, and \eqref{LT},  we explore the dynamics of entanglement, quantum memory, coherence, and entropy in the spin system. All the functions are evaluated for the final density matrix obtained in Eq. \eqref{final rho}.

\par
\begin{figure}[ht]

		\includegraphics[width=0.2\textwidth, height=125px]{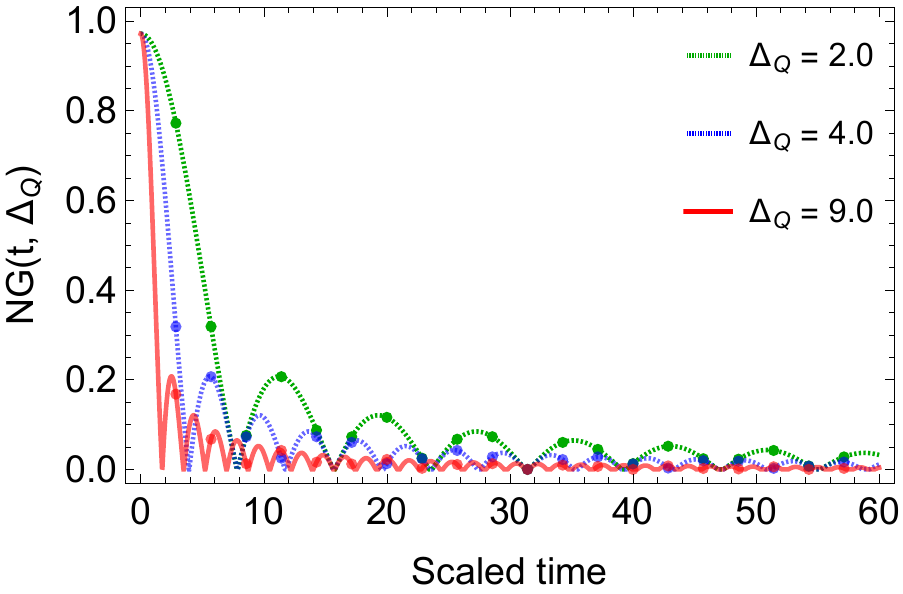}
		\put(-120,05){($ a $)}
		\includegraphics[width=0.2\textwidth, height=125px]{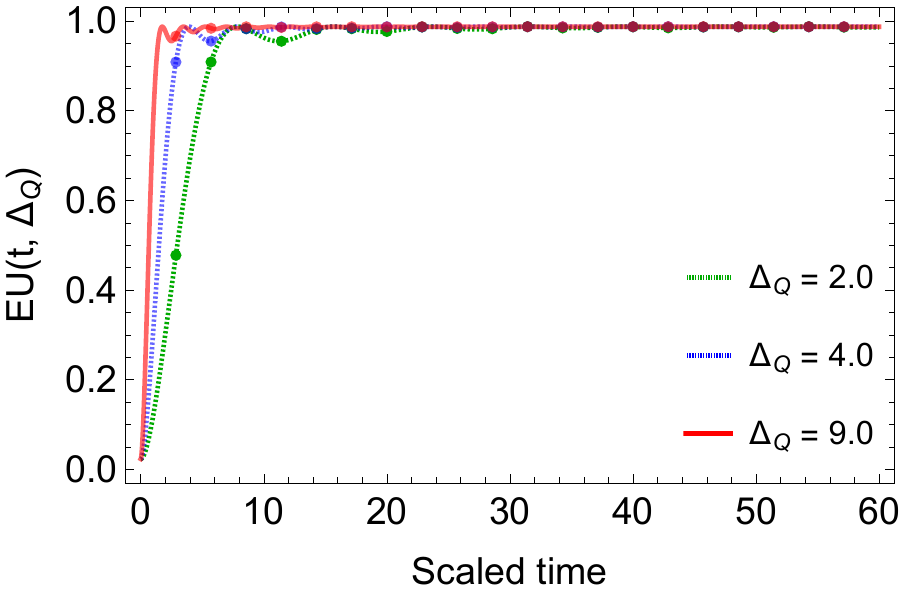}
		\put(-120,05){($ b $)}
		\includegraphics[width=0.2\textwidth, height=125px]{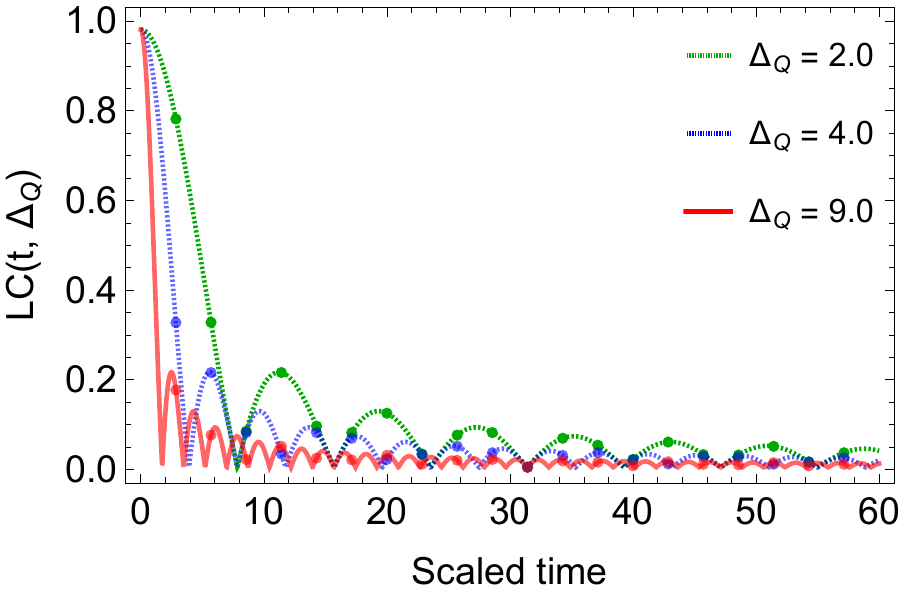}
		\put(-120,05){($ c $)}
		\includegraphics[width=0.2\textwidth, height=125px]{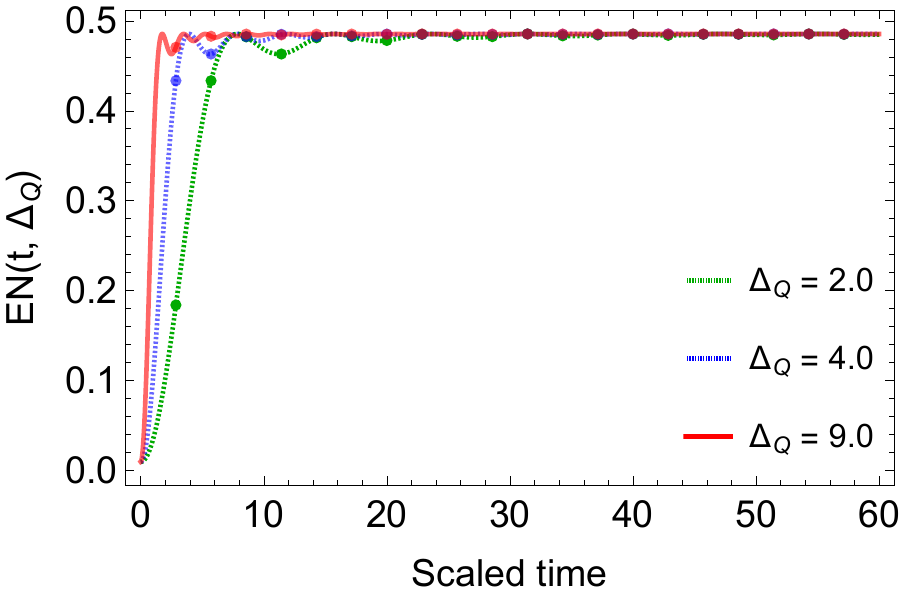}
		\put(-120,5){($ d $)}
	\caption{Dynamics of negativity (a), entropic uncertainty (b), $\ell_1$-norm coherence (c), and linear entropy (d) as functions of static noise disorder parameter $\Delta_Q$ against time in a two-spin system influenced by an external TMCC. For all the plots, we have set $\lambda =0.1$, $K_z=5$, and $J/T/D_z/J_z/\Delta_z/B/J/=1$. \label{fig1-Delta_m}}
\end{figure}
In Fig. \ref{fig1-Delta_m}, we analyze the dynamics of entanglement, entropic uncertainty, coherence, and entropy in a two-spin state when coupled with an external TMCC. Initially, the state remains maximally entangled and coherent as seen that $NG=LC=1$. The depicted entropic uncertainty and entropy in the state remain zero initially, as $EU=EN=0$. After the onset, the action of the joint TMCC appears and as a result, entanglement as well as coherence becomes easily lost. In agreement, the $EU$ and $EN$ remained increasing functions of the entropic uncertainty and disorder in the system. The overall dynamical maps of the spin state correlations remained non-Markovian, however, the degree of non-Markovianity shown by each measure differs. For example, the $NG$ and $LC$ functions show a large number of entanglement and coherence revivals. This suggests that the two-qubit system and coupled fields strongly support information exchange between them. On the other hand, the non-Markovian behavior shown by the $EU$ and $EN$ function remains weaker. Therefore, $NG$ and $LC$ functions are more sensitive than the $EU$ and $EN$ functions and record the least exchange of attributes between the state and the field. Besides, the $NG$ and $LC$ functions show anti-correlation with $EU$ and $EN$ functions, as both pairs evolve in opposite directions to each other. Furthermore, for the increasing values of the disorder parameter $\Delta_Q$, the entanglement, and coherence remain fragile and become easily lost in the state. On the other hand, the uncertainty between observables of the particles and associated mixedness disorder in the state is enhanced by the increasing strength of $\Delta_Q$.  However, it is noticeable that the rate of revivals is directly dependent upon $\Delta_Q$. The phenomenon of sudden death and birth of entanglement occurs repeatedly. Finally, entanglement and coherence functions seem completely decay with time while the uncertainty and entropy functions achieve a final higher saturation level.

\par
\begin{figure}[ht]
		\includegraphics[width=0.2\textwidth, height=125px]{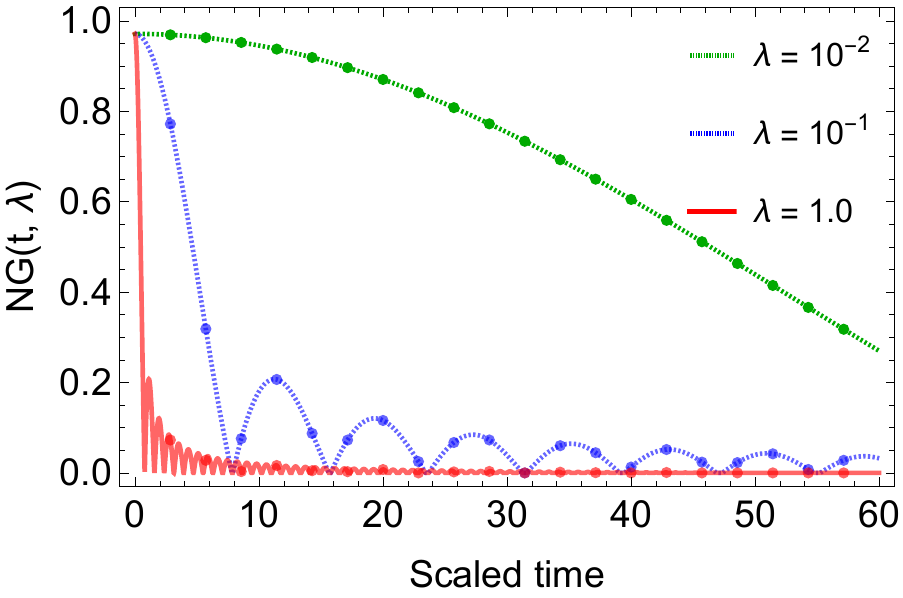}
		\put(-120,05){($ a $)}
		\includegraphics[width=0.2\textwidth, height=125px]{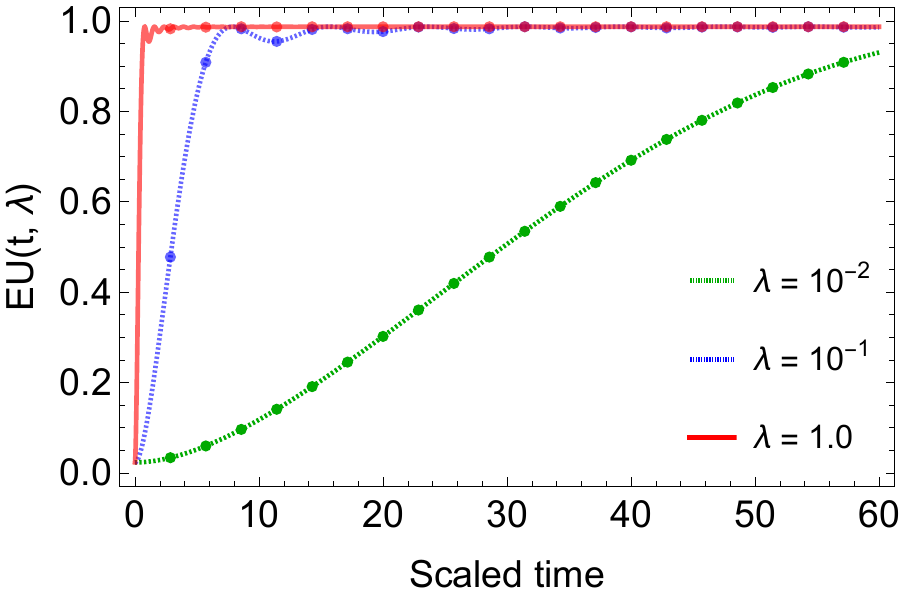}
		\put(-120,05){($ b $)}
		\includegraphics[width=0.2\textwidth, height=125px]{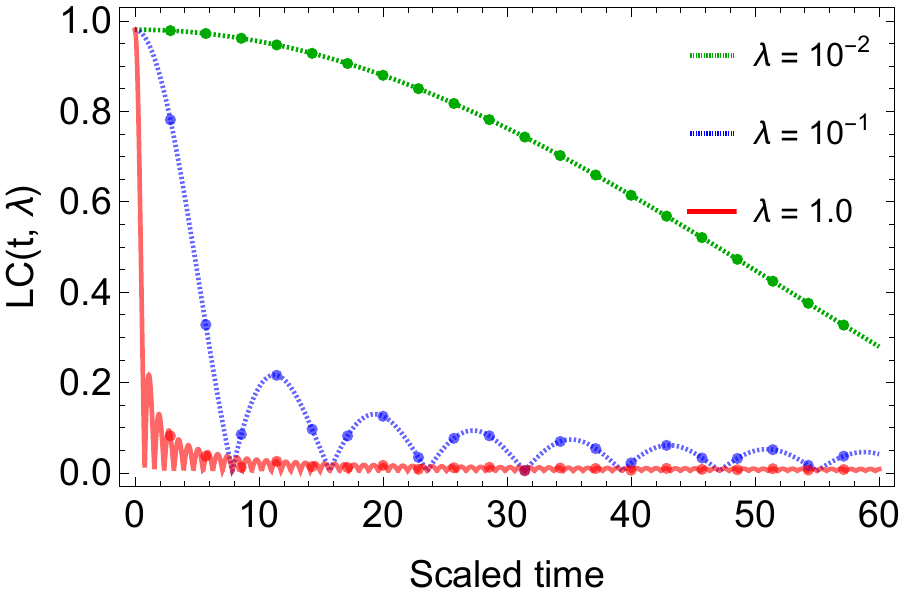}
		\put(-120,05){($ c $)}
		\includegraphics[width=0.2\textwidth, height=125px]{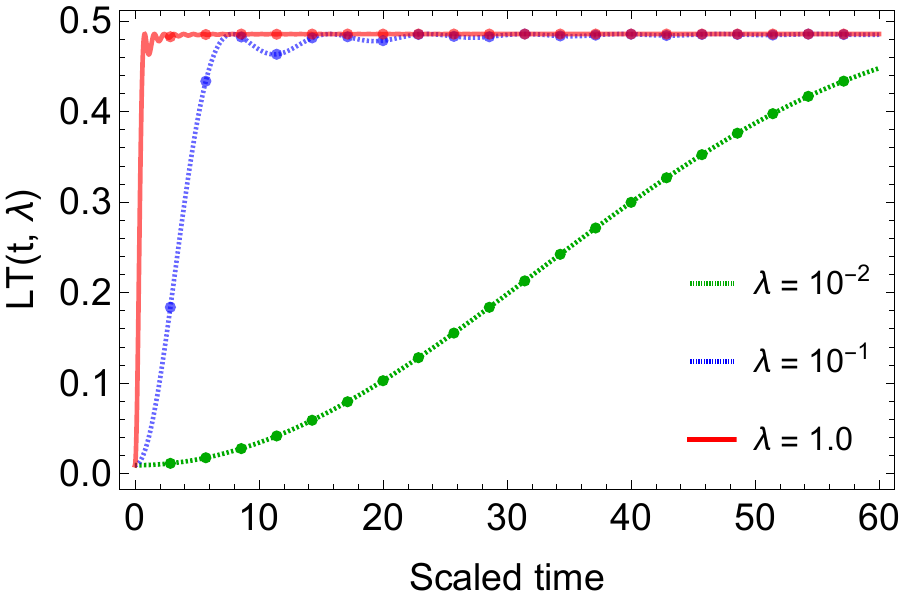}
		\put(-120,5){($ d $)}
	\caption{Dynamics of negativity (a), entropic uncertainty (b), $\ell_1$-norm coherence (c), and linear entropy (d) as functions of classical field's coupling strength $\lambda$ against time in a two-spin state influenced by an external TMCC. For all the plots, we have set $\Delta_Q=2$, $K_z=5$, and $J/T/D_z/J_z/\Delta_z/B/J=1$. \label{fig2-lambda}}
\end{figure}
In Fig. \ref{fig2-lambda}, the impact of weak as well as strong coupling strength $\lambda$ of the classical channel on the dynamics of entanglement, coherence, entropic uncertainty, and entropy disorder in a two-spin system when exposed to an external magnetic field is studied. Initially, the two-qubit spin system remains maximally entangled and coherent as the functions are $NG=LC=1$. On the contrary, the $EU=EN=0$ suggests that the system is free of entropic uncertainty and mixedness. As the interaction between the external TMCC and the two-qubit spin system starts, the initial maximal correlations in the state decrease. The speed of decay of entanglement and coherence is regulated by the coupling strength of the classical field. The strong coupling of the spin state with the classical environment results in a quicker decay of correlations and vice versa. Besides, the non-Markovian behavior of the classical fields is highly enhanced in the strong coupling regimes while becoming negligible at the weaker coupling strength end.  Likewise, the uncertainty and entropy disorder increases with higher speeds as $\lambda$ increases. Finally, the speed of entropic uncertainty and entropy is directly related to the decay rate of the entanglement and coherence. Therefore, showing an anti-correlation between the two pairs of phenomena. The phenomenon of sudden death and birth of entanglement varies with the varying strengths of coupling strength regimes of the classical field. Compared to Fig. \ref{fig1-Delta_m}, the decay observed in the current case has the least values. Hence, quantum correlation decay can be highly controlled by tuning the coupling intensity between the classical field and the two-qubit state. However, the fact remains consistent that both the parameters, namely the disorder $\Delta_Q$ and coupling parameter $\lambda$ regulated the rate of sudden death and birth of the entanglement and coherence. Finally, for the higher $\lambda$ values, the entanglement and coherence become completely lost, hence, suggesting maximum entropy disorder and absolute separability in the state.

\par

\begin{figure}[ht]

		\includegraphics[width=0.2\textwidth, height=125px]{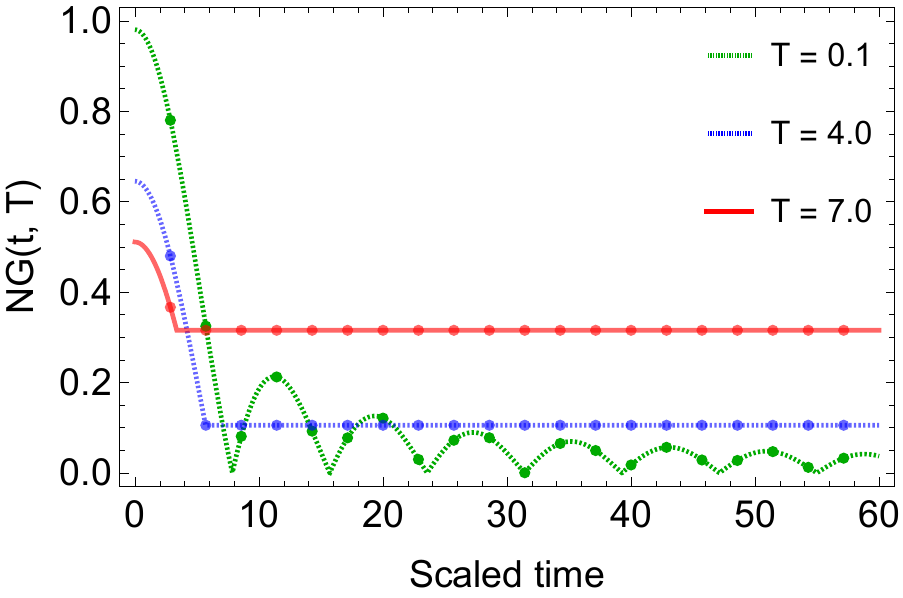}
		\put(-120,05){($ a $)}
		\includegraphics[width=0.2\textwidth, height=125px]{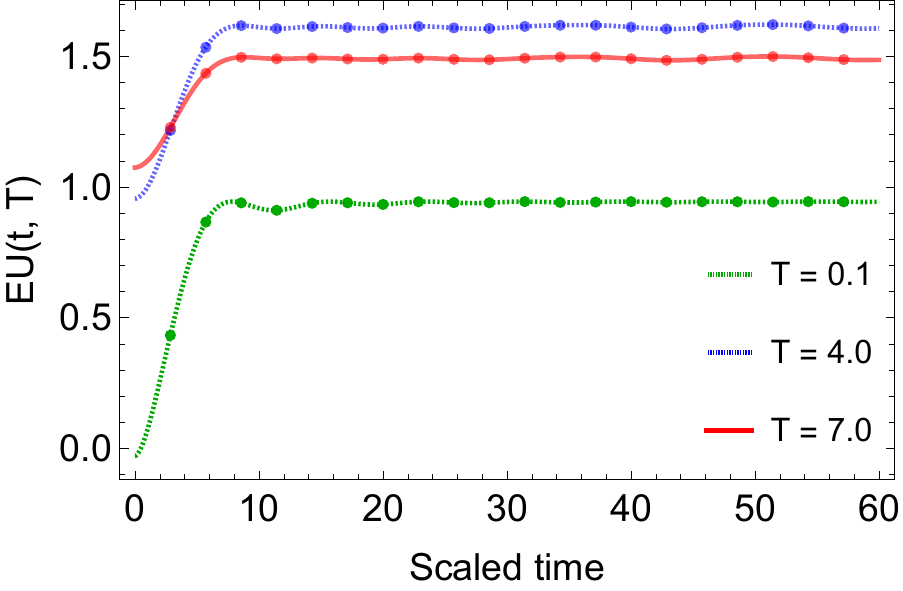}
		\put(-120,05){($ b $)}
		\includegraphics[width=0.2\textwidth, height=125px]{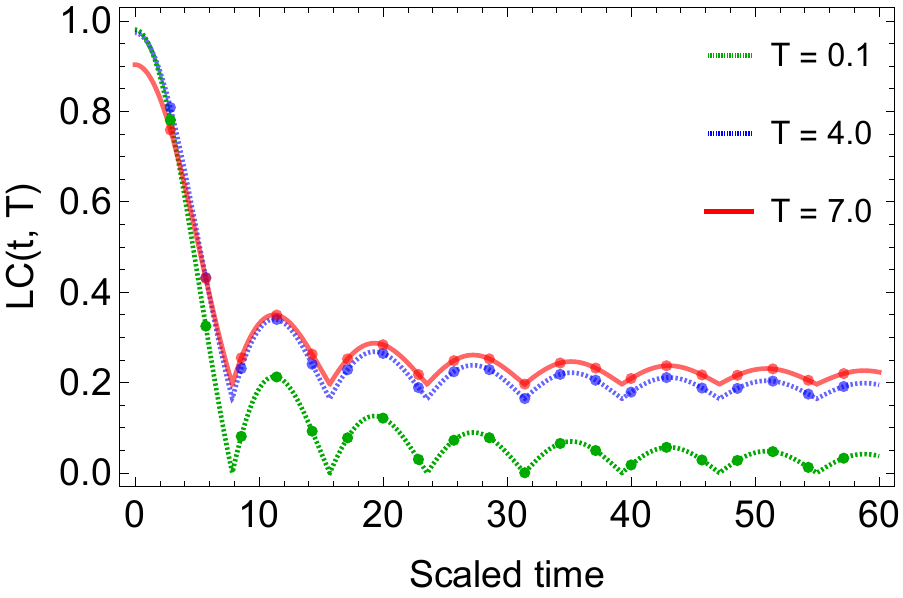}
		\put(-120,05){($ c $)}
		\includegraphics[width=0.2\textwidth, height=125px]{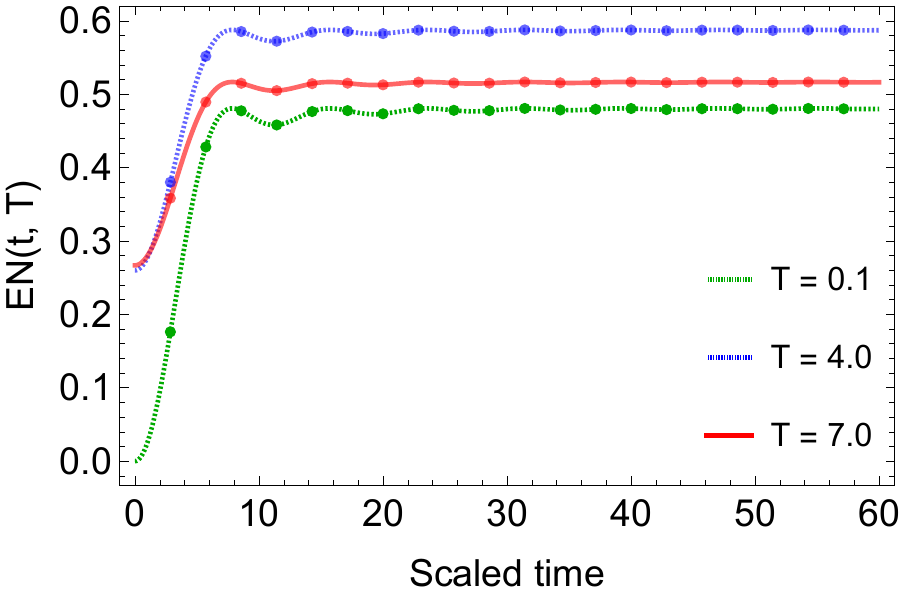}
		\put(-120,5){($ d $)}
	\caption{Dynamics of negativity (a), entropic uncertainty (b), $\ell_1$-norm coherence (c), and linear entropy (d) as functions of temperature $T$ against time in a two-spin system influenced by an external TMCC. For all the plots, we have set $\Delta_Q=2$, $\lambda=0.1$, $K_z=5$ and  $J/D_z/J_z/\Delta_z/B/J=1$. \label{fig3-Temp}}
\end{figure}
The temperature influence on the dynamics of entanglement, coherence, entropic uncertainty, and entropy is analyzed in a system of two-qubit spin state when subjected jointly to TMCC in Fig. \ref{fig3-Temp}. The influence is taken into account when the temperature of coupled fields is assumed at different fixed higher and lower values. The difference between the dynamics of the entanglement, coherence, uncertainty, and disorder functions is significant. As seen for the higher temperature values $(T=7)$, the initial entanglement and coherence values decrease while the original entropic uncertainty and mixedness in the state increase. Hence, contradicting the characteristic properties of disorder parameter $\Delta_Q$ (Fig. \ref{fig1-Delta_m}) and $\lambda$ (Fig. \ref{fig2-lambda}) which do not influence the initial values of the inclusive functions. Besides this,  the dynamical maps of entanglement comprise repeated sudden death and birth revivals for the lower temperature values ($T=0.1$). On the contrary, the entanglement decays exponentially with time for the higher temperature values. The coherence function $LC$ deviates from the entanglement $NG$ function and shows revivals of coherence, even at the higher temperature values. Hence, suggesting the strengthened nature of coherence in the spin state compared to the associated entanglement. Unlike the cases in Fig. \ref{fig1-Delta_m} and \ref{fig2-lambda}, entanglement and coherence remain preserved for longer interval of time in the current case, especially, for the higher temperature values. It is interesting to note that for the higher temperature values, the initial values of the entanglement and coherence functions decrease but remain largely preserved for the latter interval of time. Likewise, the entropic uncertainty and entropy functions are initially directly affected by the temperature, however, for the higher temperature values, the uncertainty and mixedness in the state become reduced at the latter notes of time. However, for the lower entropic uncertainty and mixedness, one must keep the temperature at minimum values. In comparison, the $NG$ function's revival character seems very susceptible to the increasing values of temperature and becomes easily dissipated.
\par
\begin{figure}[ht]
		\includegraphics[width=0.2\textwidth, height=125px]{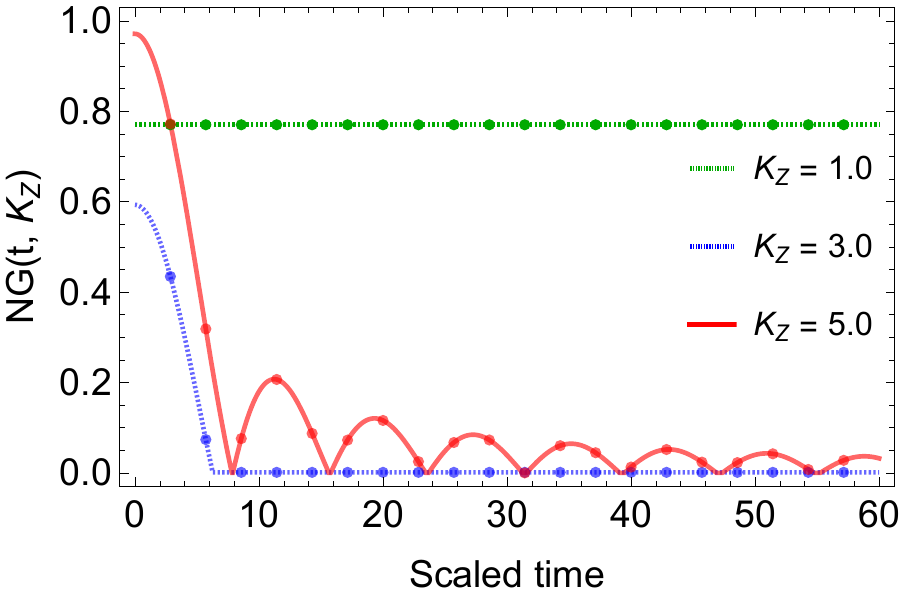}
		\put(-120,05){($ a $)}
		\includegraphics[width=0.2\textwidth, height=125px]{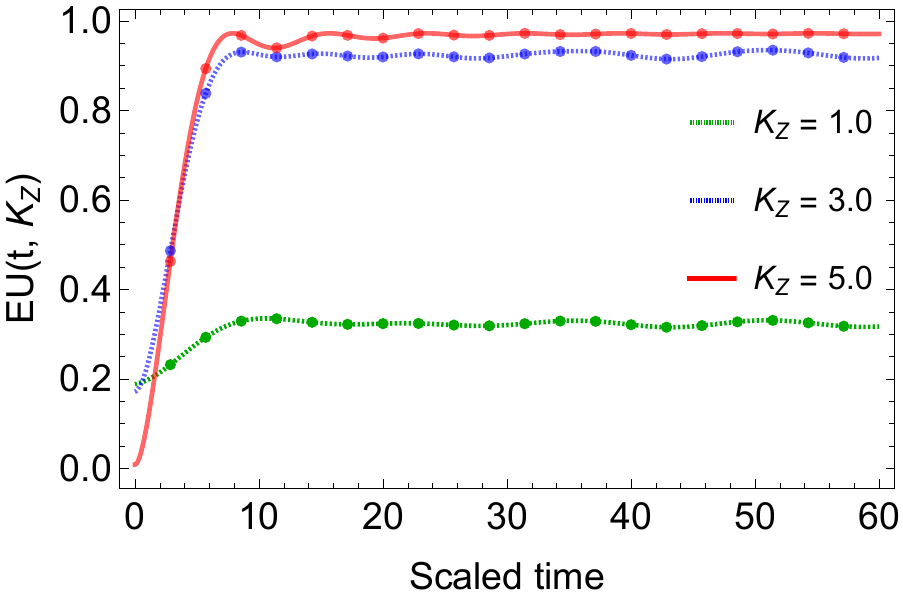}
		\put(-120,05){($ b $)}
		\includegraphics[width=0.2\textwidth, height=125px]{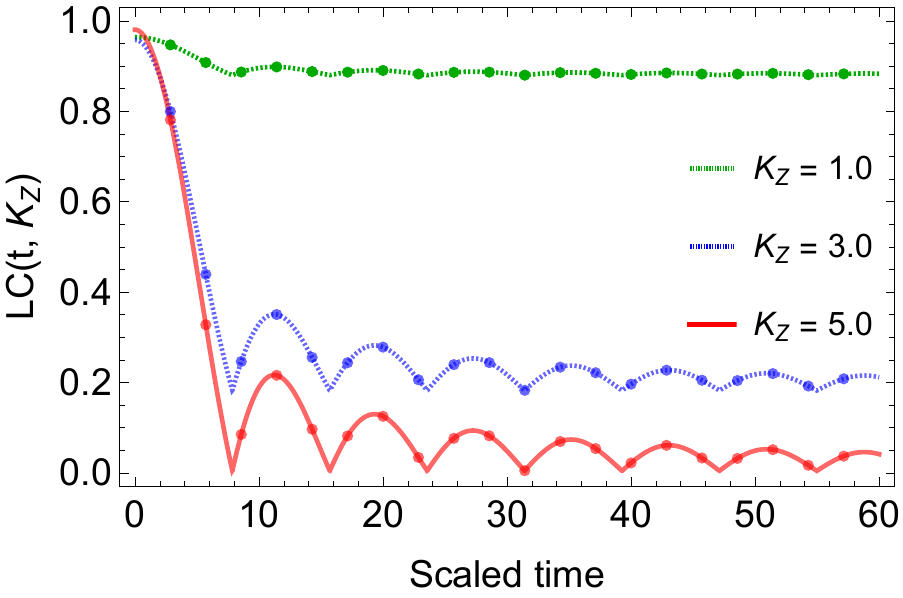}
		\put(-120,05){($ c $)}
		\includegraphics[width=0.2\textwidth, height=125px]{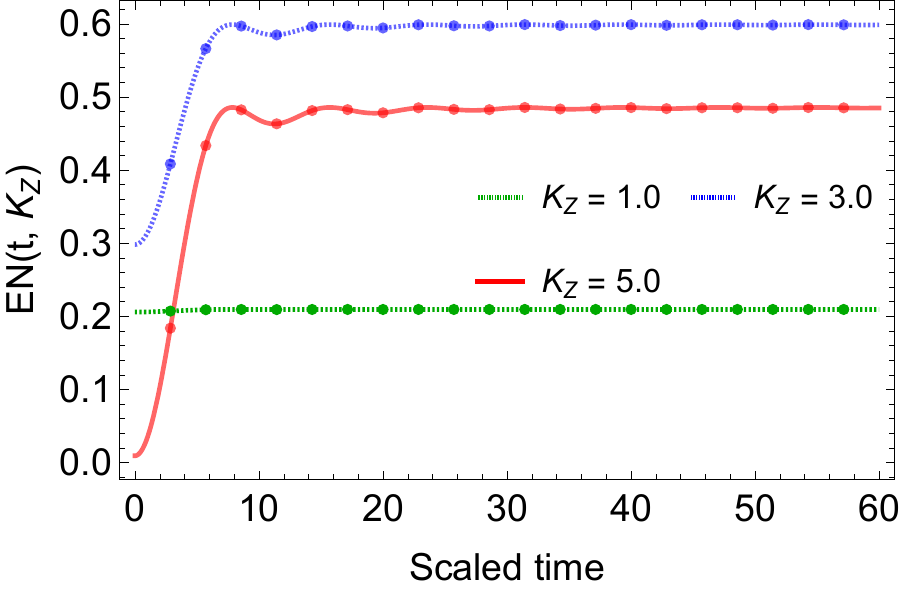}
		\put(-120,5){($ d $)}
\caption{Dynamics of negativity (a), entropic uncertainty (b), $\ell_1$-norm coherence (c), and linear entropy (d) as functions of KSEA interaction along $z$-axis $K_z$ against time in a two-spin state influenced by an external TMCC. For all the plots, we have set $\Delta_Q=2$, $\lambda=0.1$,  and $J/D_z/J_z/\Delta_z/B/J=1$. \label{fig4-KSEA}}
\end{figure}
Fig. \ref{fig4-KSEA} discloses the dynamics of entanglement, coherence, entropic uncertainty, and mixedness in a two-qubit spin system coupled with TMCC. In particular, the impact of KSEA interaction $(K_z)$ fixed to different strengths, is evaluated on the dynamics of the system. The dynamical maps of $NG$, $EU$, $LC$, and $EN$ functions seem increasingly different from those observed in Figs. \ref{fig1-Delta_m}-\ref{fig3-Temp}. As can be seen that a higher degree of entanglement and coherence preservation limit is achieved for lower KSEA interaction strength $K_z=1.0$. In close connection, a lower limit of entropic uncertainty and mixedness in the two-qubit state has been observed at $K_z=1.0$. However, for the increasing strength of the KSEA interaction ($K_z=5$), the entanglement, as well as the coherence function, suffers a greater decay. This contradicts most of the previous findings obtained in Refs. \cite{31, 68, 69, 70} where for the increasing strength of $K_z$, quantum correlations in the state becomes more preserved. However, it is noticeable that the increasing KSEA interaction strength improves the revival character of the $NG$, $EU$, $LC$, and $EN$ functions. Particularly, for the $K_z=5.0$, the entanglement and coherence functions start facing sudden deaths and births with decreasing amplitudes. Besides, entanglement for $K_z=3.0$ dissipates completely and only agrees with the entropy disorder in the state. The coherence function on the other hand remains non-zero for $K_z=3.0$ and agrees with the entropic uncertainty function. As can be seen that for $K_z=3.0$, the $EU$ function shows minimal uncertainty between the observables of the state that that seen at $K_z=5.0$.  Hence, the KSEA interaction of the external magnetic field most likely differently affects different quantum criteria. The revivals in entanglement and coherence functions suggest strong information exchange between the two-qubit spin state and coupled fields.
Therefore, with the increasing KSEA interaction, one can reverse the information lost from the system and the conversion of free-classical states into resourceful non-local states is feasible. Finally, entanglement and coherence seem indefinitely preserved in the two-qubit state for the lower $K_z$ values which may greatly benefit the quantum information processing protocols. Besides, the decay rates of entanglement and coherence are found proportional to the increasing speed of entropic uncertainty and mixedness in the state, therefore, showing an inverse relation between them. In comparison, the $LC$ followed by $NG$ functions catch a greater number of revivals compared to the $EU$ and $EN$ functions. This means that $NG$ and $LC$ are highly sensitive to the stimuli caused by the external TMCC.
\par
\begin{figure}[ht]

		\includegraphics[width=0.2\textwidth, height=125px]{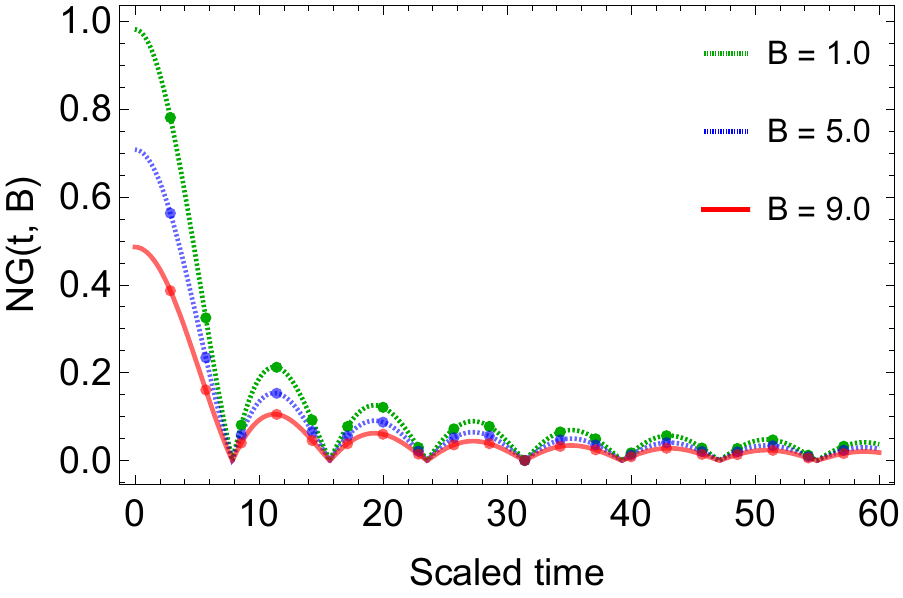}
		\put(-120,05){($ a $)}
		\includegraphics[width=0.2\textwidth, height=125px]{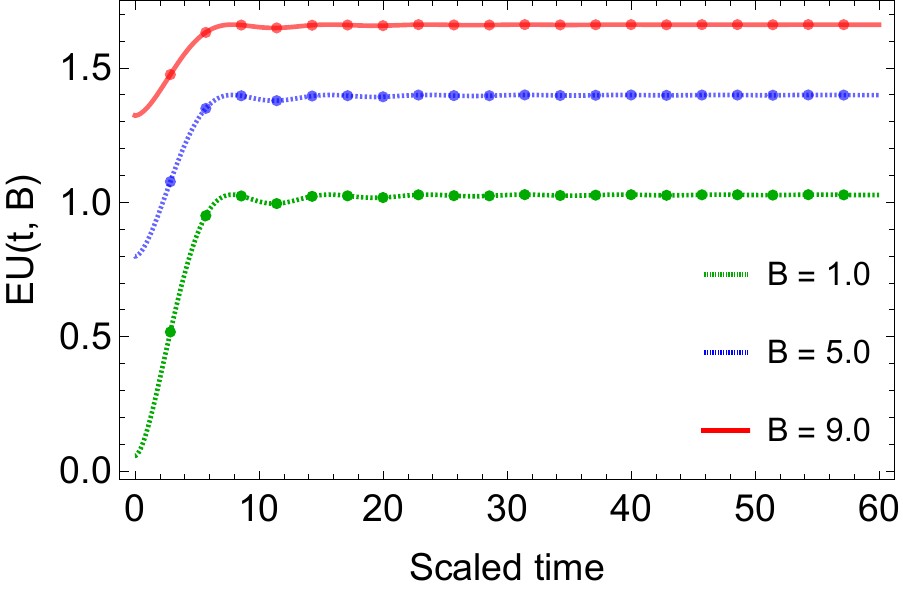}
		\put(-120,05){($ b $)}
		\includegraphics[width=0.2\textwidth, height=125px]{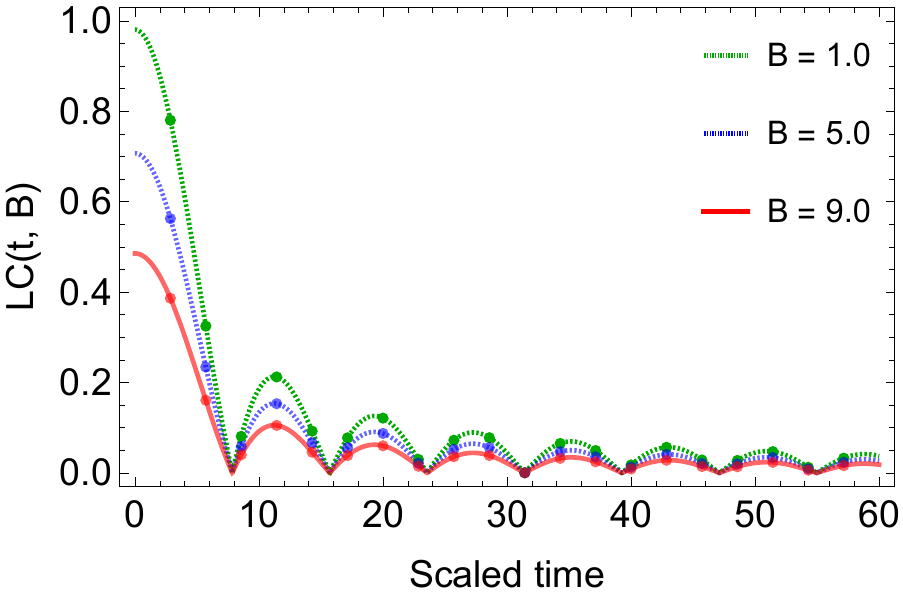}
		\put(-120,05){($ c $)}
		\includegraphics[width=0.2\textwidth, height=125px]{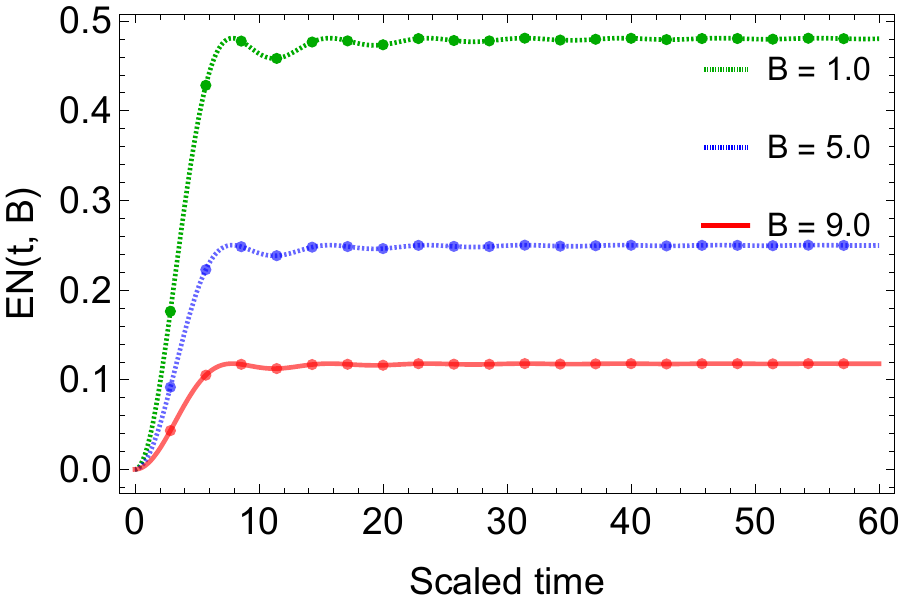}
		\put(-120,5){($ d $)}
\caption{Dynamics of negativity (a), entropic uncertainty (b), $\ell_1$-norm coherence (c), and linear entropy (d) as functions of magnetic field strength $B$ against time in a two-spin state influenced by an external TMCC. For all the plots, we have set $\Delta_Q=2$, $\lambda=0.1$, $K_z=5$, and $J/D_z/J_z/\Delta_z/J=1$. \label{fig5-magnetic}}
\end{figure}
In Fig. \ref{fig5-magnetic}, we probe the influence of different fixed values of magnetic field strength parameter $B$ on the time evolution of the two-spin system when connected with TMCC. At the onset, the state preserves different values of initial entanglement and coherence. For the higher magnetic field strengths, $B=9.0$, the initially encoded entanglement and coherence decrease while the entropic uncertainty and mixedness in the state increase. Therefore, the external magnetic  field negatively affects the preservation of quantum correlations in the spin state.  Besides the initial level, the time evolution of the entanglement and coherence functions is also negatively affected by the magnetic field. As seen for the higher magnetic field strength $B=9.0$, entanglement and coherence revive at a lower level while the entropic uncertainty and mixedness functions reach a higher saturation level. On the contrary, for the lower magnetic field strength $B=1.0$, the entanglement and coherence functions seem more preserved and the reverse can be seen for the entropic uncertainty mixedness in the state. The revival rate, on the other hand, is not disturbed by the different strengths of the external magnetic field and remains the same. However, for the lower $B$ values, the revivals in entanglement and coherence get a higher amplitude, therefore, predicting the larger information flow between the two qubits and coupled fields. In comparison, the impact of magnetic field parameter $B$ matches with that of the disorder parameter $\Delta_Q$ and coupling constant $\lambda$ (especially, the higher $\lambda$ values) of the classical field illustrated in Figs. \ref{fig1-Delta_m} and \ref{fig2-lambda}, respectively. As can be seen that in the mentioned cases, the entanglement functions quickly decay while the emergence of uncertainty disorder in the state occurs faster. Therefore, for the optimal longer and greater degree of quantum correlations in the two-qubit state, one must tune the external magnetic field to the least values. Moreover, in agreement with Figs. \ref{fig1-Delta_m}-\ref{fig4-KSEA}, we find an anti-correlation between the growing $EU$ and $EN$ functions with the decaying $NG$ and $LC$ functions. Finally, the decreasing amplitudes of the revivals in entanglement and coherence show that the state at the final notes of time will become fully separable with a higher degree of uncertainty and disorder, depending upon the strength of the external magnetic field.
\par
\begin{figure}[ht]

		\includegraphics[width=0.2\textwidth, height=125px]{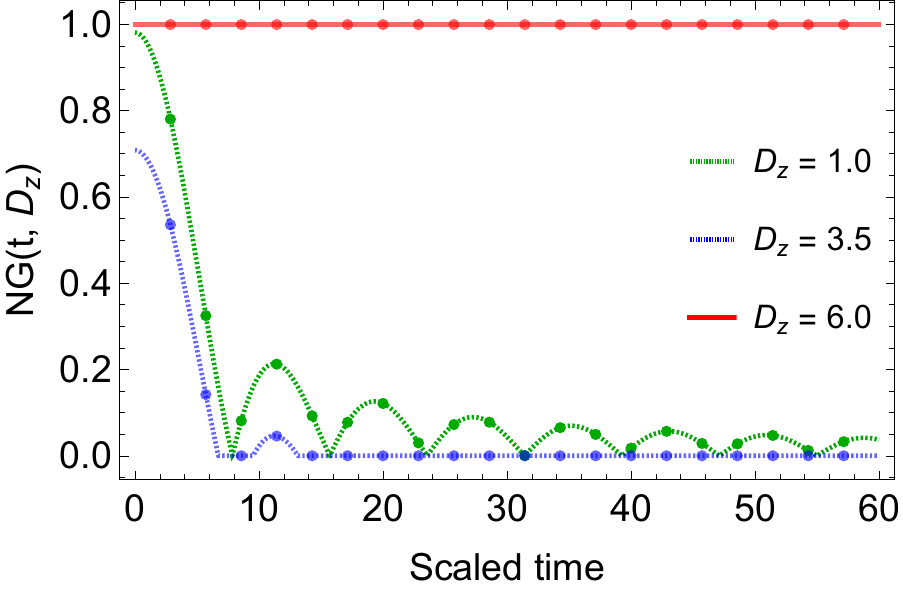}
		\put(-120,05){($ a $)}
		\includegraphics[width=0.2\textwidth, height=125px]{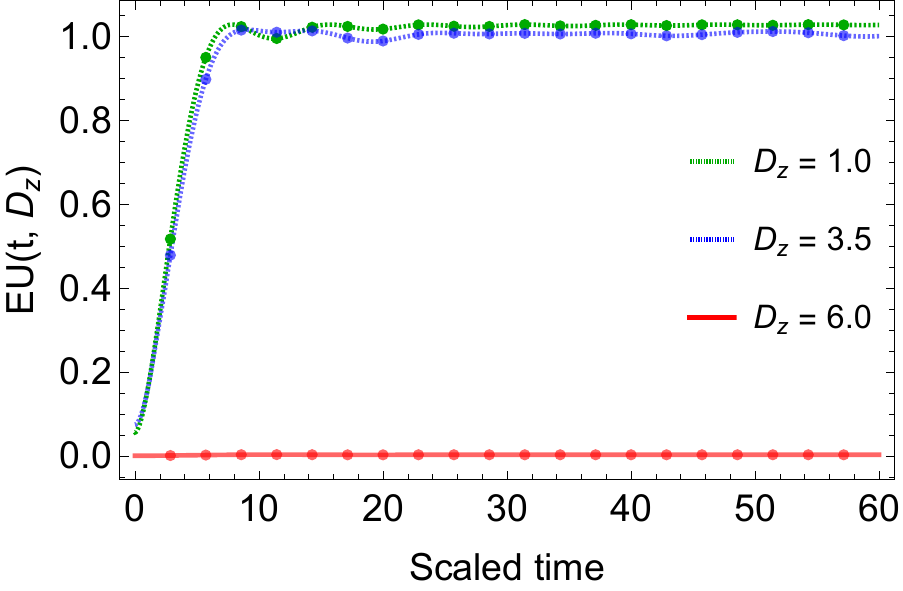}
		\put(-120,05){($ b $)}
		\includegraphics[width=0.2\textwidth, height=125px]{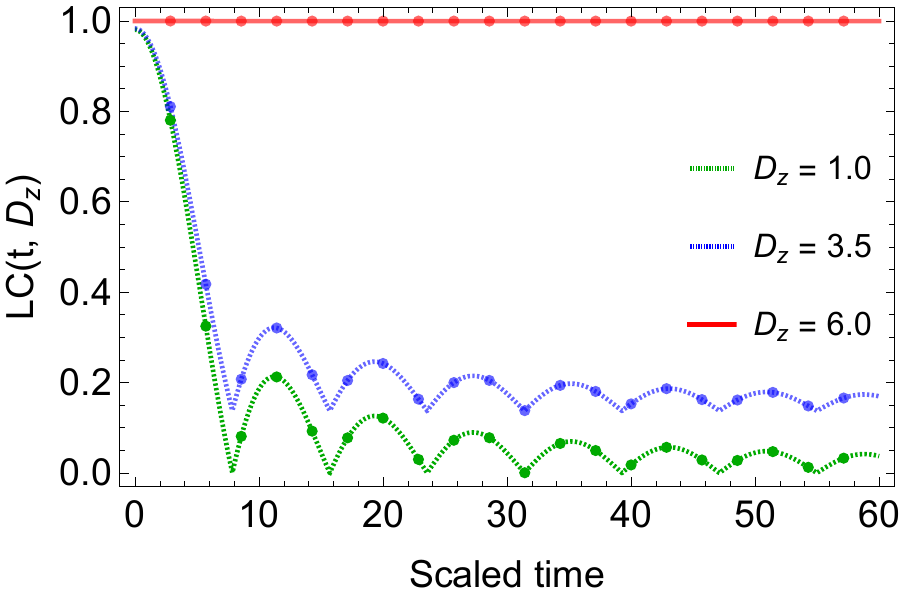}
		\put(-120,05){($ c $)}
		\includegraphics[width=0.2\textwidth, height=125px]{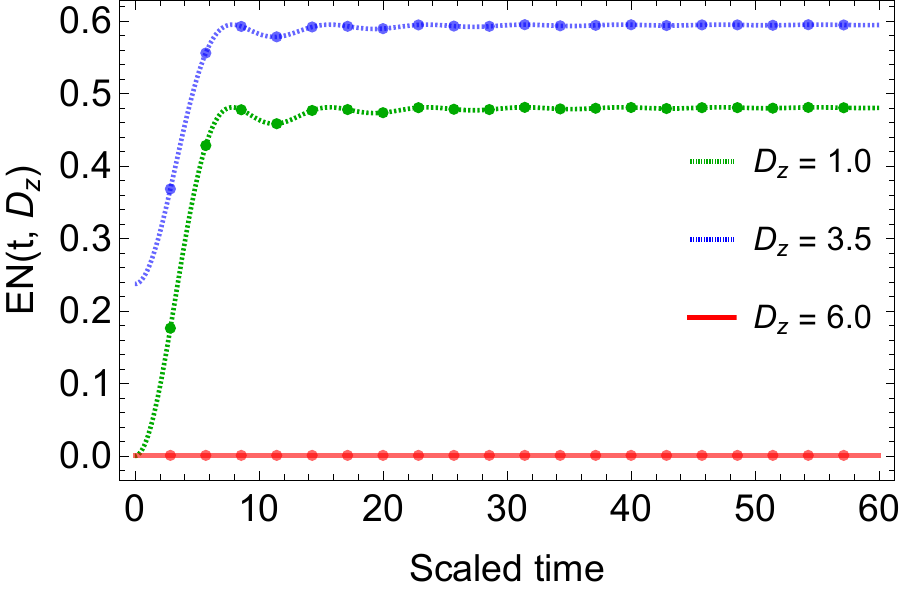}
		\put(-120,5){($ d $)}
\caption{Dynamics of negativity (a), entropic uncertainty (b), $\ell_1$-norm coherence (c), and linear entropy (d) as functions of DM interaction strength along $z$-axis $D_z$ against time in a two-spin state influenced by an external TMCC. For all the plots, we have set $\Delta_Q=2$, $\lambda=0.1$, $K_z=5$, and $J/D_z/\Delta_z/B/J=1$. \label{fig6-DM}}
\end{figure}
When coupled with TMCC configuration, the impact of different fixed values of the DM interaction strength $D_z$ on the dynamics of entanglement, coherence, entropic uncertainty, and entropy functions in a two-qubit spin state is investigated in Fig. \ref{fig6-DM}. The role of $D_z$ parameter in the dynamics of entanglement, coherence, uncertainty, and mixedness is found to be contradicting all the cases studied in Figs. \ref{fig1-Delta_m}-\ref{fig5-magnetic}. In all the previous figures, the parameters associated with the classical as well as the magnetic field seem negatively affecting the initially encoded quantum correlations and positively accelerated the uncertainty and mixedness function. However, the opposite occurs in the current case where for the increasing strength of $D_z$, entanglement and coherence functions get enhanced while the uncertainty and entropy functions get suppressed. Precisely, the two-qubit state for the initial as well as for the later interval of time remains maximally entangled and coherent for the $D_z=6.0$. In agreement, for $D_z=6.0$, the state remains indefinitely free of the uncertainty and mixedness disorder which is interesting. However, for $D_z<5$, the agreement between the entanglement and coherence functions vanishes where the prior one becomes non-maximal while the latter one remains maximal. It is also interesting that for $D_z=3.5$, entanglement shows a quick drop and seems completely dissipating while the coherence yet remains preserved with a non-zero value. The initial, as well as the later dynamical map of entanglement functions, seems in close connection with the entropy function. On the other hand, the $LC$ function and the associated dynamical outlooks seem similar to that obtained for the uncertainty in the state. As seen for different values of $D_z$, the initial level of entanglement and entropy differs. However, for a similar situation, the initial levels of coherence and uncertainty correspond to each other and both functions start from maximal coherence and zero uncertainty points, respectively. The revival characteristic of the entanglement and coherence appears enhanced for the lower $D_z$ values and so the opposite. Finally, by regulating the $D_z$ to higher values, the state may be kept maximally entangled and coherent while free from uncertainty and entropy disorder. While detuning of the $D_z$ may cause the initial as well as the latter preserved levels of quantum correlations to be decreased while enhancement in the uncertainty and entropy in the state may occur. Besides this, the fact remains consistent that there is an anti-correlation between the entanglement and coherence pair with the uncertainty and entropy pair.
\par
\begin{figure}[ht]

		\includegraphics[width=0.2\textwidth, height=125px]{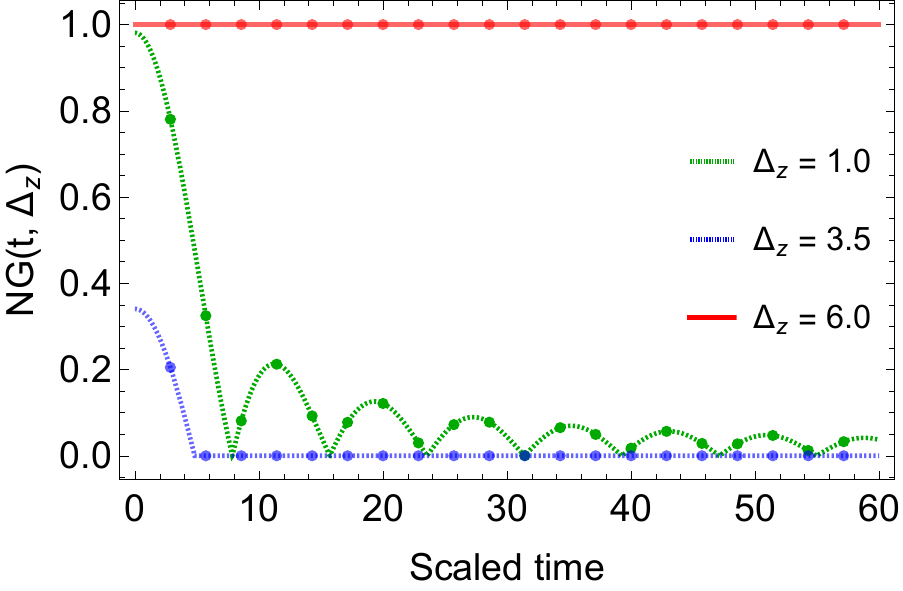}
		\put(-120,05){($ a $)}
		\includegraphics[width=0.2\textwidth, height=125px]{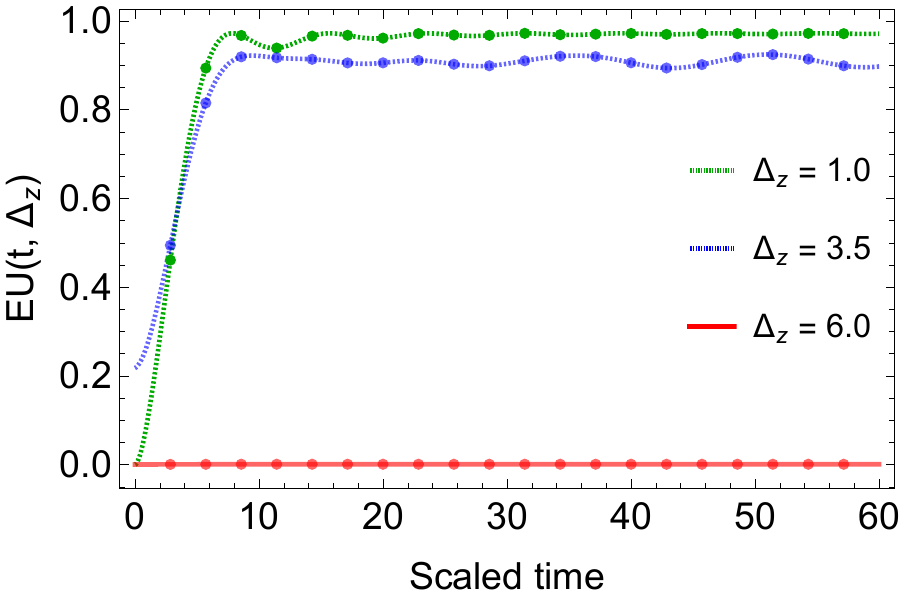}
		\put(-120,05){($ b $)}
		\includegraphics[width=0.2\textwidth, height=125px]{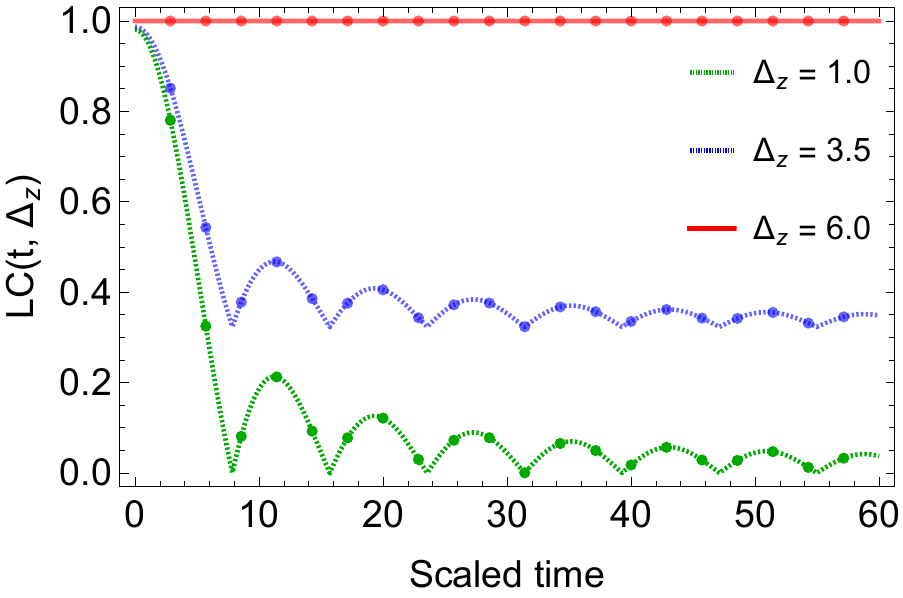}
		\put(-120,05){($ c $)}
		\includegraphics[width=0.2\textwidth, height=125px]{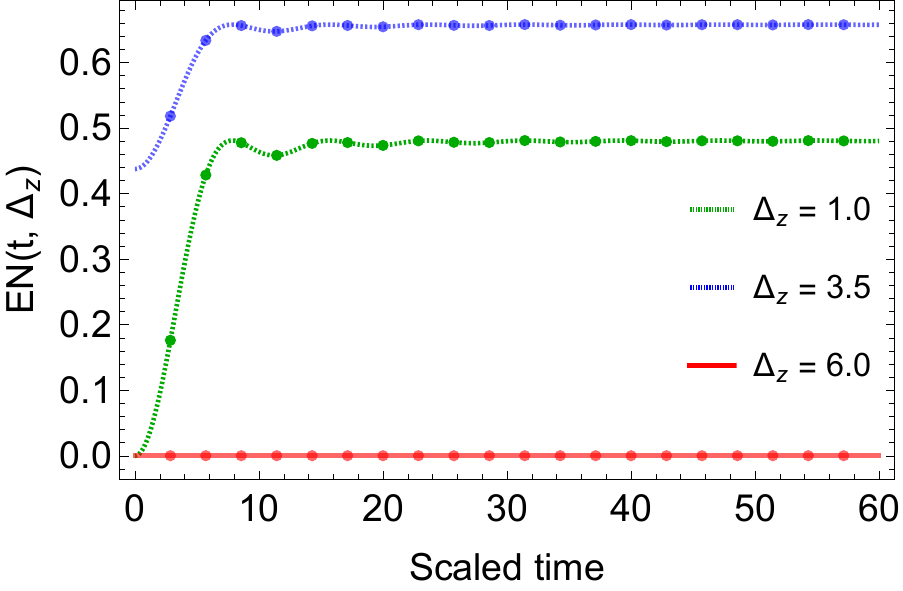}
		\put(-120,5){($ d $)}
\caption{Dynamics of negativity (a), entropic uncertainty (b), $\ell_1$-norm coherence (c), and linear entropy (d) as functions of the symmetric exchange spin-spin interaction strength in $z$-direction along $\Delta_z$ against time in a two-spin state influenced by an external TMCC. For all the plots, we have set $\Delta_Q=2$, $\lambda=0.1$, $K_z=5$, and $J/D_z/B/J=1$. \label{fig7-anisotropy}}
\end{figure}
In Fig. \ref{fig7-anisotropy}, we address the dynamics of entanglement, coherence, entropic uncertainty, and entropy mixedness in a two-spin system system when exposed to an external TMCC. The influence of different fixed values of the anisotropy factor of the two-spin system systems on the dynamics of the two-qubit correlations is obtained in detail. The dynamical outlook obtained for different values of $\Delta_z$ seems in agreement with that obtained in Fig. \ref{fig6-DM} for different fixed values of the DM-interaction. However, the results obtained contradict the results obtained in Figs. \ref{fig1-Delta_m}-\ref{fig5-magnetic} where strong entanglement and coherence decay has been detected. For the increasing anisotropy strengths, entanglement and coherence remained highly preserved. For instance, see the slopes of entanglement and coherence at $\Delta_z=6.0$ where the state remains maximally entangled and coherent indefinitely. In close connection, for $\Delta_z=6.0$, the $EU$ and $EN$ functions remain zero, hence, predicting zero-entropic uncertainty and disorder in the spin state. It is interesting to note that for $\Delta_z=3.5$, coherence is improved while entanglement has been witnessed to decay completely in a short time and the same has been also witnessed in Fig. \ref{fig6-DM} against $D_z$. Entanglement and coherence show repeated sudden death and birth revivals at $\Delta_z=1.0$, therefore, predicting strong information exchange between the two-qubit state and coupled fields. The entropy function $EN$ agrees with entanglement function $NG$ (where entanglement completely decays) and shows that for $\Delta_z=3.5$, entropy in the system becomes maximum. On the other hand, the $LC$ and $EU$ functions agree and show that minimum coherence decay and uncertainty rise occurs for $\Delta_z=3.5$ when compared to $\Delta_z=1.0$. The entanglement and coherence functions stay in inverse relation with the entropic uncertainty/entropy functions, as can be seen, that both grow in opposite directions.

\par
\begin{figure}[!h]
	\begin{center}
		\includegraphics[width=0.35\textwidth, height=175px]{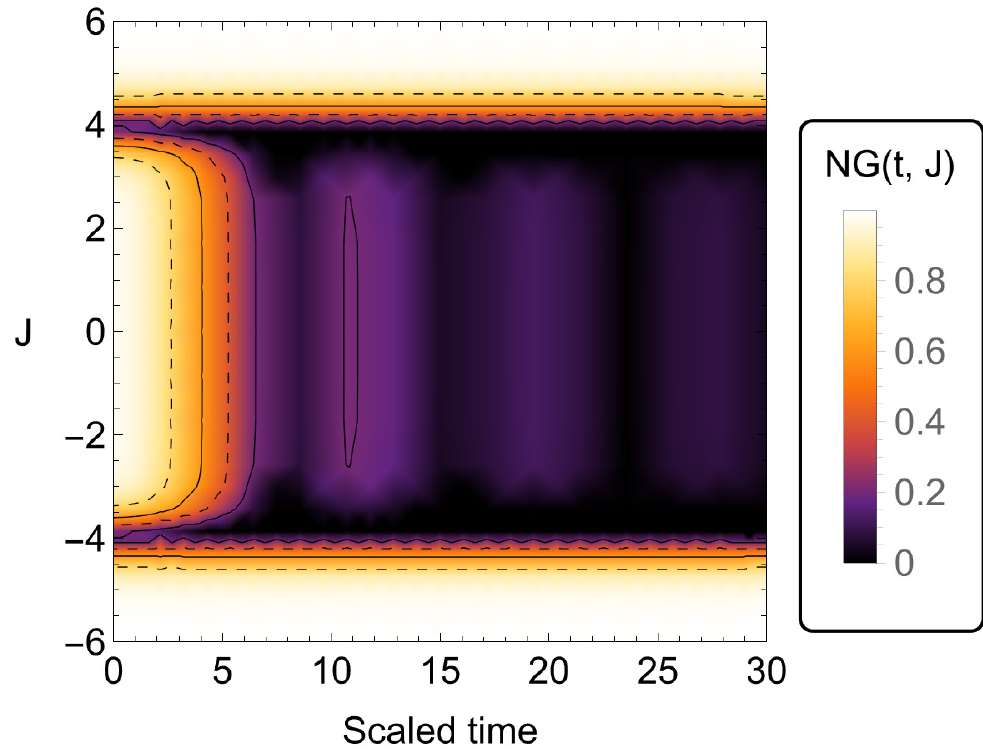}
		\put(-200,180){($ a $)}\
		\includegraphics[width=0.35\textwidth, height=175px]{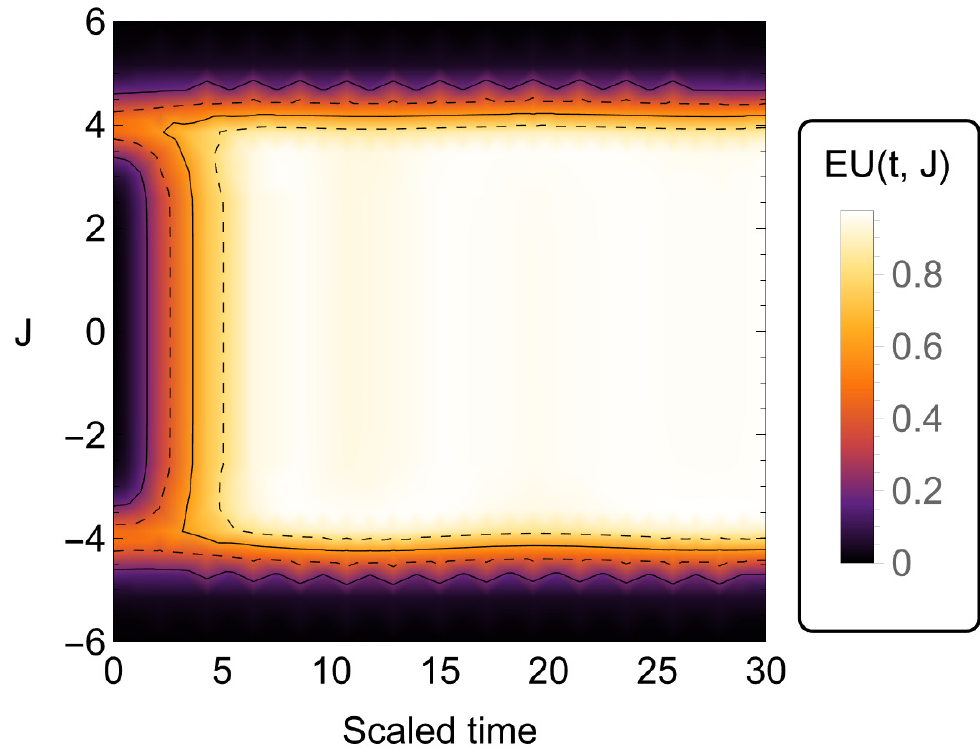}
		\put(-200,180){($ b $)}\ \\
		\includegraphics[width=0.35\textwidth, height=175px]{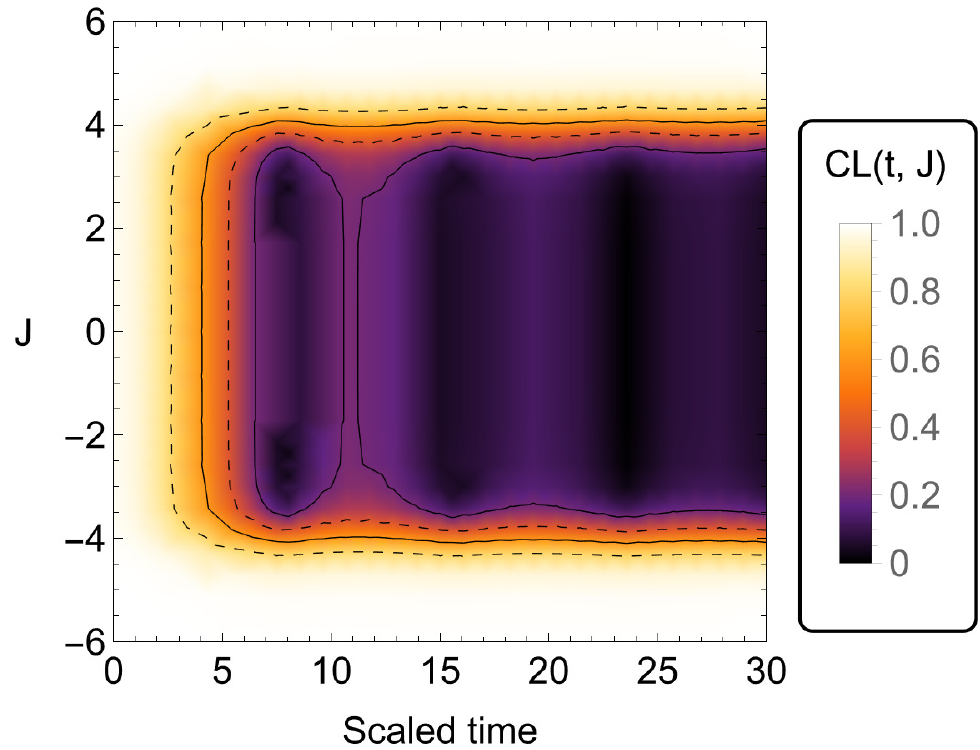}
		\put(-200,180){($ c $)}\
		\includegraphics[width=0.35\textwidth, height=175px]{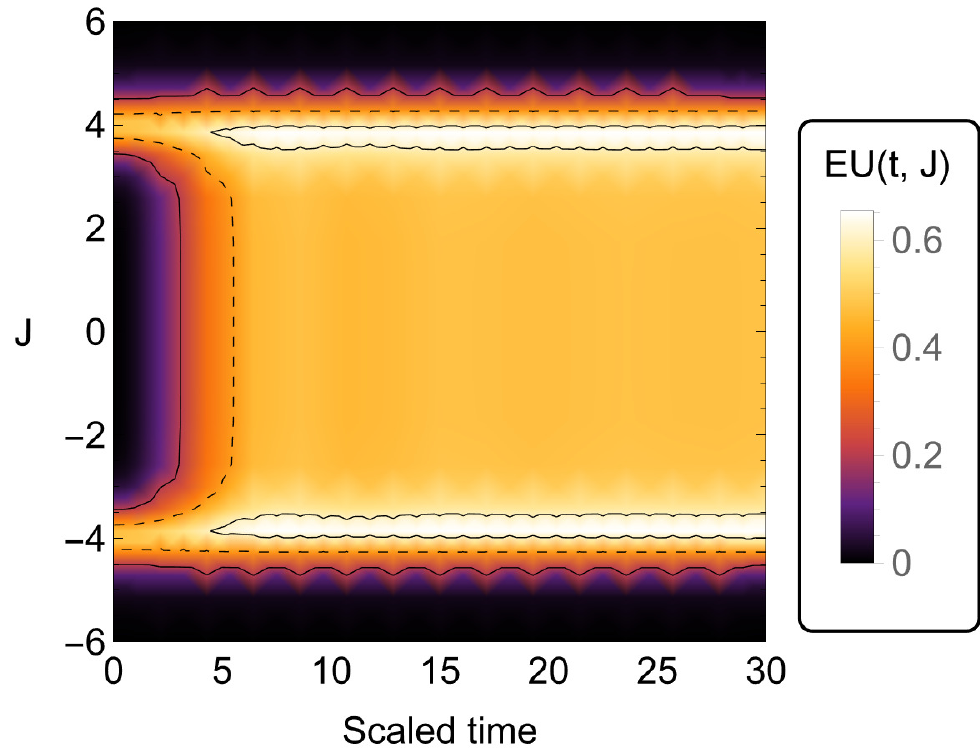}
		\put(-200,180){($ d $)}
		\end{center}
\caption{Dynamics of negativity (a), entropic uncertainty (b), $\ell_1$-norm coherence (c), and linear entropy (d) as functions of the Heisenberg exchange interaction strength $J$ against time in a two-spin state influenced by an external TMCC. For all the plots, we have set $\Delta_Q=2$, $\lambda=0.1$, $K_z=5$, $T=0.5$ and $J/D_z/B/J=1$.\label{fig8-J}}
\end{figure}
In Fig. \ref{fig8-J}, we investigate the impact of Heisenberg exchange interaction parameter $J$ on the preservation of entanglement/coherence and generation of entropic uncertainty and entropy disorder in the spin state exposed to the TMCC. We have demonstrated the results for the ferromagnetic ($J<0$) as well as for the antiferromagnetic ($J>0$) regimes. The impact of both the ferromagnetic and antiferromagnetic regimes have been found similarly affecting the generation of entanglement and suppression of the entropic uncertainty and disorder in the state. As can be seen that for both the positive and negative regimes of the Heisenberg exchange interaction, the entanglement and coherence functions gradually generate while the entropic uncertainty and entropy functions decay with time. Notice that for certain critical ranges, such as $\pm 3.5 <J < \pm 4.5$, the state seems initially partially entangled while having non-zero uncertainty and mixedness. Interestingly, coherence in the state is unaffected by this range and the state remains maximally entangled and coherent. Therefore, suggesting the strengthened nature of coherence compared to the entanglement. In the range $+4.5 <J < -4.5$, the dynamics of the state exhibit either zero- or partially preserved entanglement and coherence regimes at the latter intervals of time. However, for the range, $\pm 4.5 <J =\pm 6.0$, the state remains maximally entangled and coherent while encountering no decay. In agreement, for the forenamed region, the state remains completely free of uncertainty and entropy. In comparison, the coherence function exhibits a larger number of revivals of coherence compared to the other inclusive functions. As seen that all other functions exhibit a lower number of revivals or completely dissipate after a short interval of time.
\par
\subsection{Fidelity of the state}
Finally, we give a brief time to cover how much the state becomes distinguished from the one originally assumed. In this regard, the notion of fidelity can be used to determine the discrimination between any two given states. Let one of the states is  $\rho_{st}(t, T)$ given in Eq. \eqref{final rho} and the other one be any arbitrary state $\sigma$. Further, the $\sigma $ state is assumed into two catteries: the first given in the Eq. \eqref{state} namely, $\rho(0, T) $. In the second case, let us consider a maximally entangled state $\psi=\frac{1}{\sqrt{2}}(\ket{00}+\ket{00})$, therefore, the initial density matrix of the state becomes as $\rho_0=\ket{\psi}\bra{\psi}$. Therefore, $\sigma \in \{\rho(0, T), \rho_0)\}$. Using such a comparison would lead us to clarify the distance between the resultant, initial thermal, or maximally entangled two-qubit state.\\
In this case, fidelity for the two matrices $\rho_{st}(t, T)$  and $\sigma$ can be written as
\begin{equation}
FID_X=Tr[\rho_{st}(t, T)\sigma]+2\sqrt{\det[\rho_{st}(t, T)] \det[\sigma]},
\end{equation}
where fidelity between state  $\rho_{st}(t, T)$ and  $\rho(0, T)$ becomes as
\begin{equation}
FID_1=X1+X2+X3
\end{equation}
where
\begin{align*}
X1=&\rho_{11}^2+\rho_{22}^2+\rho_{33}^2+\rho_{44}^2+2\rho_{23}\rho_{23}^*,&\\
X2=&\frac{\rho_{14} \rho_{14}^* \sin (2 \Delta_Q \lambda  t) \cos (4 \Delta_o \lambda  t)}{\Delta_Q \lambda  t},&\\
X3=&\sqrt{\frac{\left(| \rho_{14}| ^2- \rho_{11}  \rho_{44}\right) \left( \rho_{22}  \rho_{33}- \rho_{23}  \rho_{23}^*\right)^2 \left( \rho_{14} \rho_{14}^* \sin ^2(2 \Delta_Q \lambda  t)-4 \Delta_Q^2 \lambda ^2  \rho_{11}  \rho_{44} t^2\right)}{\Delta_Q^2 \lambda ^2 t^2}}.&
\end{align*}
Besides, fidelity between the $\rho_{st}(t, T)$  and maximally entangled state has the form
\begin{equation}
FID_{2}=\frac{1}{4} \left(2 (\rho_{11}+\rho_{44})+\frac{e^{-4 i \Delta_o \lambda  t} \sin (2\Delta_Q \lambda  t) \left(\rho_{14}+\rho_{14}^* e^{8 i \Delta_o \lambda  t}\right)}{\Delta_Q \lambda  t}\right)
\end{equation}
\begin{figure}[!h]
	\begin{center}
		\includegraphics[width=0.35\textwidth, height=175px]{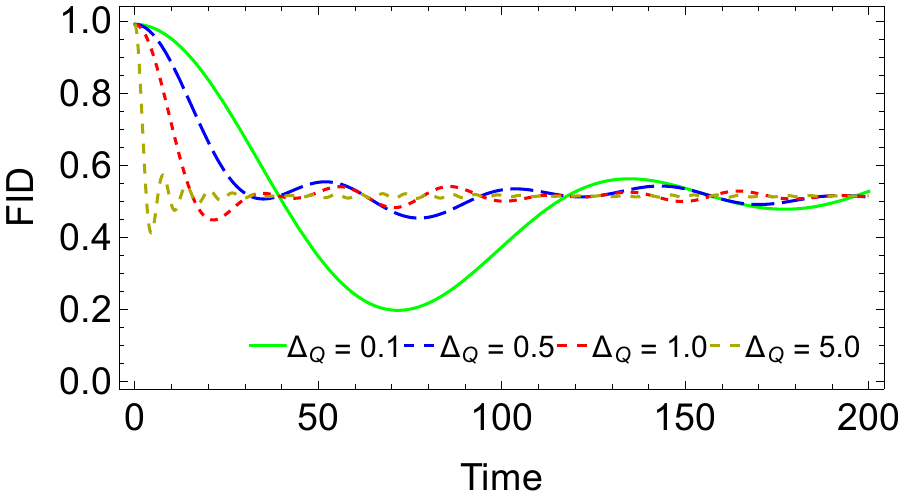}
		\put(-200,180){($ a $)}\ 
		\includegraphics[width=0.35\textwidth, height=175px]{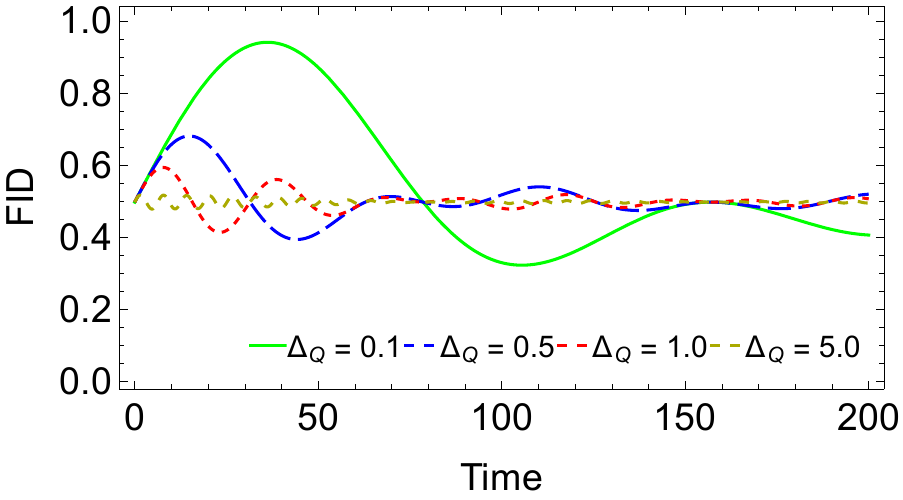}
		\end{center}
\caption{(a)Dynamics of fidelity between states  $\rho_{st}(t, T)$ and  $\rho(0, T)$ for the two-qubit case when exposed to the hybrid channel against various strengths of classical dephasing while setting $\lambda=0.1$, $K_z=5$, $T=0.5$ and $J/D_z/B/J=1$. (b) Same as (a) but for $\rho_{st}(t, T)$ and  $\rho_0$.\label{fig10}}
\end{figure}
In Fig. \ref{fig10}, the fidelity dynamics for between the states  $\rho_{st}(t, T)-\rho(0, T)$ ($FID_1$) and  $\rho_{st}(t, T)-\rho_0$ ($FID_2$) is presented against various static noisy dephasing parameter. This would enable us to predict how much the resultant state deviates from the originally considered state under strong and weak dephasing limits. In Fig. \ref{fig10}(a), the two states seem completely comparable, however, with time they become distinguished more and more. However, the fidelity loss rate is mostly concentrated on the dephasing strength introduced in the system. For example, for $\Delta_Q=5.0$, the fidelity loss rate is enough quicker, and so the opposite can be seen at the weak dephasing limits. However, for the weak dephasing limits, the minimum achieved by the green slopes is deeper than that seen for the strong dephasing limits. This shows that the states become more distinguishable at a specific duration. Besides, for the stronger dephasing strengths, the slopes show quicker revivals as compared to that at the weak dephasing end. After a long time, all the slopes finally seem to achieve a similar saturation level, hence, predicting, a similar amount of fidelity even for the different $\Delta_Q$ values. In the second case Fig. \ref{fig10}(b), initially, the state remains more distinguishable, however. the fidelity increases between the state $\rho_{st}(t, T)$ and $\rho_0$. The rate of fidelity increase remains higher for the weak dephasing strengths and decreased for the higher dephasing strengths. However, with time, the slopes for the different values of $\Delta_Q$ seem to accumulate the same saturation level. This shows that there would remain a constant amount of distinguishability between the maximally entangled and our considered state $\rho_{st}(t, T)$ in the current given conditions.

\par
Furthermore, in the current work, we have found that entropic uncertainty and entropic disorder negatively affect the degree of quantum correlations in the state. As seen that when the entropic uncertainty and disorder rise, entanglement and coherence decay. It is noticeable that the rate of entropic uncertainty and mixedness remains higher than the decay rate of entanglement and coherence. This suggests that the entropic uncertainty and disorder in the two-qubit spin states lead while the opposite involved functions lag. Hence, illustrating the rise in uncertainty and entropy as the major causes of the loss of quantum correlations in the spin systems. Besides, we believe that the current configuration can be successfully employed for the transmission of quantum information in associated protocols. As seen that this configuration can be readily exploitable to induce longer quantum correlations preservation. Moreover, the cases of implementation of individual classical channels have been found damaging for the quantum correlations preservation compared to when the classical channel is employed jointly with the external magnetic field \cite{49, 50, 51, 63}. In addition, the current results also appraise that the same configuration with individual external magnetic fields and thermal fields exhibit lesser and shorter preserved quantum correlations compared to when is jointly employed with classical field \cite{31}.
\section{Experimental feasibility}
Any configuration, without its experimental nature, would be assumed not to be beneficial. For this reason, we give some experimental prospects for the current studied configuration. We find that classical channels have been experimentally well adopted with quantum channels previously for information processing, for example, see Refs. \cite{ee1, ee2, ee3, ee4}. In Ref. \cite{ee1}, by keeping a spacing of $1.6$nm between the local and non-local channels while tuning the frequency to $200$GHz at $-24$dBm, the authors disclosed a resourceful joint channel. They found that the total coexistence power in single mode fiber (SMF), where the secret key rate (SKR) was recorded to drop by 73 $\%$. Besides this, at $0$dBm total coexistence power in hollow core nested antiresonant nodeless (HC-NAN) fiber ($250$ times greater power saving than that achieved for SMF), the SKR remains more preserved. The co-deployment of a discrete variable-quantum key distribution channel with $8 \times 200$Gbps classical channels was resourcefully demonstrated using a $2$km long HC-NAN fiber with a high speedy transmission of $1.6$Tbps \cite{ee2}. Here, the authors claimed the coexistence of classical-quantum channels are recorder to have a reduced decay and even with more power saving (nearly 40 times) compared to the individual utilization of quantum channels. In relevance to our study, the authors in Ref. \cite{ee3} experimentally probed counter-propagation of local-quantum channels over a $1$km-long $7-$core, multicore fiber and found that they show high tolerance to the noise compared to the individual use of the counterparts, hence, showing agreement with our results. The authors in Ref. \cite{ee4} realized a quantum–classical channel over a $7$-core multicore fiber, based on space division multiplexing with the highest launch power of $25$dBm. By looking into the above studies, it can be readily deduced that classical channels can be coined together with quantum channels such as the ones studied here. Moreover, we propose that if the current configuration is utilized by applying certain procedures, then it might turn out to be a resourceful way for the practical transmission of quantum information, quantum devices, and associated quantum operations.

\section{Conclusion}\label{conclusion}
This study disclosed the characterization of a hybrid channel comprising thermal, magnetic, and classical parts characterized by various quantum correlations strengthening and weakening characteristics. The case of a two-qubit for the sake of simplicity is owned to investigate the time evolution of entanglement, coherence, entropic uncertainty, and entropy disorder under the considered channel. various quantum tools are used to characterize the parameters of the configuration, and the resultant dynamical maps of the assumed quantum criteria have been explicitly studied. Finally, we provided the optimal parameter setting which best suits quantum correlations preservation and entropic uncertainty as well as disorder suppression under the action of the hybrid channel.
\par
We demonstrate that the collective symmetry of the current configuration shared by the thermal, magnetic, and classical fields continues to be a crucial factor in maintaining the preservation of quantum correlations.. There exist certain aspects in this joint external field setup can be usefully used to avoid the emergence of entropic uncertainty and entropy disorder in the state. For example, in the weaker coupling regimes of the classical field, certain values of DM, spin-spin, and anisotropy interaction strengths, the state remains maximally entangled and coherent initially as well as for the latter interval of time. In close connection, for the given conditions, the state can be kept completely free of uncertainty and entropy disorder. The entanglement and coherence decay has been found completely depending upon the emergence of the entropic uncertainty and entropy disorder in the two-qubit spin state system. Certain aspects of the coupled fields, such as the coupling strengths, disorder parameter of the static noise, magnetic field strength, and temperature of the hybrid channel have been found responsible for the emergence of uncertainty and entropy as well as for the decay of the entanglement and coherence. Interestingly, previously the KSEA interaction has been found highly influential for the preservation of quantum correlations, however, the opposite has been witnessed when imposed with the current hybrid channel. Finally, we believe that the intersection of the thermal, classical, and magnetic fields along with certain state parameters remains a vital choice that shows the capacity to be controlled easily for the preservation of quantum data compared to that when they are considered individually.
\section*{Author contributions}
Conceptualization, Atta ur Rahman; Data curation, Ma-Cheng Yang and Cong-Feng Qiao ; Formal analysis, Ma-Cheng Yang, S. M. Zangi and Cong-Feng Qiao ; Funding acquisition, Cong-Feng Qiao ; Investigation, Atta ur Rahman; Methodology, S. M. Zangi and Cong-Feng Qiao ; Project administration, Cong-Feng Qiao ; Resources, Atta ur Rahman and Cong-Feng Qiao ; Software, Atta ur Rahman; Supervision, Cong-Feng Qiao ; Validation, Ma-Cheng Yang, S. M. Zangi and Cong-Feng Qiao ; Visualization, Atta ur Rahman; Writing-original draft, Atta ur Rahman and Cong-Feng Qiao ; Writing-review  and editing, Atta ur Rahman.
\section*{Acknowledgments}
\noindent
This work was supported in part by the National Natural Science Foundation of China (NSFC) under the Grants 11975236, 12235008 and by the University of Chinese Academy of Sciences.

\section*{Data availability}
\noindent No datasets were generated or analyzed during the current study.

\section*{Competing interests}
\noindent The authors declare no competing interests.

\section*{ORCID iDs}
\noindent Atta ur Rahman \href{https://orcid.org/0000-0001-7058-5671}{https://orcid.org/0000-0001-7058-5671}\\
\noindent  Ma-Cheng Yang \href{https://orcid.org/---------}{https://orcid.org/---------}\\
\noindent S. M. Zangi \href{https://orcid.org/0000-0002-4601-4681}{https://orcid.org/0000-0002-4601-4681}\\
\noindent Cong-Feng Qiao \href{https://orcid.org/0000-0002-9174-7307}{https://orcid.org/0000-0002-9174-7307}\\

\end{document}